\documentclass [11 pt,epsfig]{amsart}

\usepackage{amsmath}
\usepackage{amscd}
\usepackage{amsthm}
\usepackage{amssymb}
\usepackage{amsfonts}
\usepackage{epsfig}
\usepackage{amsmath,amscd}
\input{xypic}
\numberwithin{equation}{section}

\newcommand{\ie}{{\it i.e.\/}\ }
\newcommand{\eg}{{\it e.g.\/}\ }
\newcommand{\cf}{{\it cf.\/}\ }

\newtheorem{thm}{Theorem}[section]
\newtheorem{cor}[thm]{Corollary}
\newtheorem{lem}[thm]{Lemma}
\newtheorem{prop}[thm]{Proposition}
\theoremstyle{definition}
\newtheorem{defn}[thm]{Definition}
\theoremstyle{remark}

\newtheorem{exa}[thm]{Example}
\numberwithin{equation}{section}

\newcommand{\bq}{\begin{eqnarray}}
\newcommand{\nq}{\end{eqnarray}}

\def\dirac{\mathpalette\cancel\partial}

\newcommand{\m}{M}
\newcommand{\R}{\mathbb{R}}
\newcommand{\s}{S^3}

\def\bU{{\mathbb U}}

\def\C{{\mathbb C}}

\def\N{{\mathbb N}}

\def\Q{{\mathbb Q}}
\def\Z{{\mathbb Z}}
\def\R{{\mathbb R}}

\def\cA{{\mathcal A}}
\def\cB{{\mathcal B}}
\def\cC{{\mathcal C}}
\def\cD{{\mathcal D}}

\def\cG{{\mathcal G}}
\def\cH{{\mathcal H}}

\def\cK{{\mathcal K}}

\def\cM{{\mathcal M}}

\def\cR{{\mathcal R}}
\def\cS{{\mathcal S}}

\def\cU{{\mathcal U}}

\def\cY{{\mathcal Y}}

\def\Aut{{\rm Aut}}

\def\End{{\rm End}}
\def\Ext{{\rm Ext}}

\def\Hom{{\rm Hom}}

\def\Ind{{\rm Ind}}

\def\Ker{{\rm Ker}}

\def\Spec{{\rm Spec}}

\def\Tor{{\rm Tor}}

\def\Tr{{\rm Tr}}

\def\cancel#1#2{\ooalign{$\hfil#1\mkern1mu/\hfil$\crcr$#1#2$}}
\def\dirac{\mathpalette\cancel\partial}

\begin{document}

\title[Coverings, correspondences, and NCG]{Coverings,
correspondences, and noncommutative geometry}
\author{Matilde Marcolli and Ahmad Zainy al-Yasry}
\address{Max-Planck Institute for Mathematics \\ 
Bonn, Germany}
\email{marcolli@mpim-bonn.mpg.de} 
\address{University of Baghdad \\ Iraq \\
 and  Max-Planck Institute for Mathematics \\ 
Bonn, Germany}
\email{zainy@mpim-bonn.mpg.de}

\date{2007}%
\begin{abstract}
We construct an additive category where objects are embedded graphs 
in the 3-sphere and morphisms are {\em geometric correspondences} 
given by 3-manifolds realized in different ways as branched covers of 
the 3-sphere, up to branched cover cobordisms. 
We consider dynamical systems obtained from
associated convolution algebras endowed with time
evolutions defined in terms of the underlying geometries.
We describe the relevance of our construction to the problem of 
spectral correspondences in noncommutative geometry.
\end{abstract}
\maketitle

\section {Introduction}
In this paper we construct an additive 
category whose objects are embedded graphs in the 3-sphere and where morphisms
are formal linear combinations of 3-manifolds. Our definition of
correspondences relies on the Alexander branched covering theorem \cite{Al}, which
shows that all closed oriented 3-manifolds can be realized as branched coverings 
of the 3-sphere, with branched locus an embedded (non necessarily connected)
graph. The way in which a given 3-manifold is realized 
as a branched cover is highly not unique. It is precisely
this lack of uniqueness that makes it possible to regard 3-manifolds
as correspondences. In fact, we show that, by considering a 3-manifold $\m$ realized
in two different ways as a covering of the 3-sphere as defining
a correspondence between the branch loci of the two covering maps, we
obtain a well defined associative composition of correspondences given
by the fibered product. 

An equivalence relation between
correspondences given by 4-dimensional 
cobordisms is introduced to conveniently reduce the size of the 
spaces of morphisms. We construct a 2-category where morphisms are
coverings as above and 2-morphisms are cobordisms of branched coverings. 
We discuss how to pass from embedded graphs to embedded links using the
relation of $b$-homotopy on branched coverings, which is a special case
of the cobordism relation. 

We associate to the set of correspondences with composition
a convolution algebra and we describe natural time evolutions induced
by the multiplicity of the covering maps. We prove that, when
considering correspondences modulo the equivalence relation of
cobordism, these time evolutions are generated by a Hamiltonian with
discrete spectrum and finite multiplicity of the eigenvalues. 

Similarly,  in the case of the 2-category,
we construct an algebra of functions of cobordisms, with two product structures
corresponding to the vertical and horizontal composition of 2-morphisms. 
We consider a time evolution on this algebra, which is compatible with
the vertical composition of 2-morphism given by gluing of cobordisms,
that corresponds to the Euclidean version of Hartle--Hawking
gravity. This has the effect of weighting each cobordism according to the
corresponding Einstein--Hilbert action. 
We also show that evolutions compatible with the vertical
composition of 2-morphisms can be obtained from numerical
invariants satisfying an inclusion--exclusion principle. In particular,
we construct an example based on the splitting formula for the
index of an elliptic operator of Dirac type with APS boundary
conditions, of the type that arises, for instance, in the 
linearized version of the gluing formulae for gauge theoretic moduli 
spaces on 4-manifolds. 

The fact that we have a vertical time evolution coming from an index 
theorem and suggests that time evolutions compatible with the
horizontal compositions may also be found by considering an index pairing, 
this time obtained by applying the bivariant Chern character to the KK-classes  
associated to the geometric correspondences. We outline the argument for such a
construction.  

Our category constructed using 3-manifolds as morphisms is motivated by 
the problem of developing a suitable notion of {\em spectral correspondences} 
in noncommutative geometry, outlined in the last chapter of the book
\cite{co1}. The spectral correspondences described in
\cite{co1} will be the product of a finite noncommutative geometry
by a ``manifold part''. The latter is a smooth closed oriented
3-manifold that can be seen as a correspondence in the sense
described in the present paper. We discuss the problem of extending
the construction presented here to the case of products of manifolds
by finite noncommutative spaces in the last section of the paper.

\medskip

{\bf Acknowledgment.} We thank the referee for many very useful comments 
and suggestions. The first author is partially supported by NSF-grant 
DMS-0651925. 

\smallskip
\section{Three-manifolds as correspondences}

For the moment, we only work in the PL (piecewise linear) category, with proper PL maps.
This is no serious restriction as, in the case of 3-dimensional and 4-dimensional manifolds, 
there is no obstruction in passing from the PL to the smooth category.
When we refer to embedded graphs in $\s$, we mean PL 
embeddings of 1-complexes in $\s$ with no order zero or order one vertices.
We use the notation 
\bq \label{2covers} 
G \subset E \subset \s
\stackrel{\pi_G} \longleftarrow \m \stackrel{\pi_{G'}}
\longrightarrow \s \supset E'\supset G' 
\nq 
to denote a closed 3-manifold $\m$ that is realized
in two ways as a covering of $\s$, respectively branched along 
(not necessarily connected) embedded graphs $E$ and $E'$ containing
fixed subgraphs $G$ and $G'$.

More precisely, we consider PL manifolds endowed with a combinatorial structure. 
Possibly up to passing to a subdivision of the triangulation, we
assume that the 1-complexes $E$ and $E'$ are simplicial subcomplexes of the triangulation. 
The assumption on the proper PL covering maps is that they map simplicial complexes to
simplicial complexes and that the preimage of a simplicial complex  is also a simplicial 
subcomplex of the triangulation, possibly after subdivisions.

In particular, this setting includes the case where the branch loci are knots or links. 
As an example, in the simpler case where the branch loci are knots, we consider
the case of the Poincar\'e homology sphere.

\begin{exa}
The Poincar\'e homology sphere $\m$ can be viewed as
a 5-fold covering of $\s$ branched along the trefoil $K_{2,3}$, or
as a 3-fold cover branched along the $(2,5)$ torus knot $K_{2,5}$
of also as a 2-fold cover branched along the $(3,5)$ torus knot
$K_{3,5}$. Thus we can see $\m$ as a correspondence
$K_{2,3} \subset \s \leftarrow \m \rightarrow
\s \supset K_{2,5}$, or as $K_{2,5} \subset \s \leftarrow \m \rightarrow
\s \supset K_{3,5}$, etc.
\end{exa}

\medskip
\subsection{The set of geometric correspondences}\label{CorrSetSec}

We define the set of geometric correspondences $\cC(G,G')$
between two embedded graphs $G$ and $G'$ in the following way.

\begin{defn}\label{CKKprime}
Given two embedded graphs $G$ and $G'$ in $\s$, let $\cC(G,G')$ denote the
set of 3-manifolds $\m$ that can be represented as branched covers
as in \eqref{2covers}, for some graphs $E$ and $E'$, respectively 
containing $G$ and $G'$ as subgraphs. In the case where $G=G'$, the set
$\cC(G,G)$ also contains the trivial unbranched covering 
$id: \s \to \s$. 
\end{defn}

The following observations are meant to show that, in general, the
$\cC(G,G')$ defined as above tend to be very large. 

In fact, restricting for simplicity to the case where $G$ and $G'$
are knots, we first have, as an immediate consequence of \cite{Pier},
the following simple observation.

\begin{lem}\label{covmov}
Let $\m$ be a closed 3-manifold that is realized as a branched
cover of $\s$, branched along a knot $K$. Then the manifold $\m$
belongs to $\cC(K,K')$, for all knots $K'$ that are obtained from
$K$ by the covering moves of \cite{Pier}.
\end{lem}

It follows immediately from this that the $\cC(G,G')$ can be very large. 
In fact, because of the existence of universal knots (\cf \cite{HLM}) 
we have the following result.

\begin{lem}\label{univknots}
If the branch loci are universal knots $K$ and $K'$, then $\cC(K,K')$ 
contains all closed oriented connected 3-manifolds.
\end{lem}

To avoid logical complications in dealing with the ``set'' of all
3-manifolds, we describe the $\cC(G,G')$ in terms of the following 
set of representation theoretic data. 

It is well known \cite{fox} that a
branched covering $p:\m \to \s$ is uniquely determined by the
restriction to the complement of the branch locus $E\subset \s$.
This gives an equivalent description of branched coverings in
terms of representations of the fundamental group of the complement
of the branch locus \cite{fox2}. Namely, 
assigning a branched cover $p:\m \to \s$ of order $m$ branched along
an embedded graph $E$ is the same as assigning a representation
\begin{equation}\label{reppi1L}
\sigma_E: \pi_1(\s\smallsetminus E) \to S_m ,
\end{equation}
where $S_m$ denotes the group of permutations of $m$ elements.
The representation is determined up to inner automorphisms,
hence there is no dependence on the choice of a base point
for the fundamental group in \eqref{reppi1L}. 

Thus, in terms of these representations, the spaces of morphisms $\cC(G,G')$
are identified with the set of data
\begin{equation}\label{sets}
 \cR_{G,G'} \subset \bigcup_{n,m,G\subset E, G'\subset E'} 
\Hom(\pi_1(\s\smallsetminus E),S_n)\times \Hom(\pi_1(\s\smallsetminus
E'),S_m), 
\end{equation}
where the $E,E'$ are embedded graphs, $n,m\in \N$, and
where the subset $\cR_{G,G'}$ is determined by the condition that the
pair of representations $(\sigma_1,\sigma_2)$ define the same
3-manifold.
This latter condition is equivalent, in the case where $n=m=3$ and 
where the branch loci are knots, to the knots being related by 
covering moves (colored Reidemeister moves), as in \cite{Pier}.

\medskip

Notice how the use of manifolds (or varieties)
as correspondences is common to other contexts in mathematics,
such as the geometric correspondences of KK-theory or the
correspondences based on algebraic cycles in the theory of motives. 
As in these other theories, we will later introduce a suitable 
equivalence relation on the geometric correspondences, that reduces 
the size of the sets $\cC(G,G')$.

\medskip
\subsection{Composition of correspondences}

The first step, in order to show that we can use Definition \ref{CKKprime} 
as a good notion of morphisms in a category where objects are embedded 
graphs in the 3-sphere, is to show that we have a well defined associative 
composition rule
\begin{equation}\label{composition}
\circ: \cC(G,G')\times \cC(G',G'') \to \cC(G,G'').
\end{equation}

\begin{defn}\label{comp2}
Let $\m \in \cC(G,G')$ and $\tilde\m \in \cC(G',G'')$ be closed 
oriented PL 3-manifolds with proper PL branched covering maps 
\bq \label{MMprimeL} \begin{array}{c}
G\subset E \subset \s
\stackrel{\pi_G} \longleftarrow \m \stackrel{\pi_1}
\longrightarrow \s \supset E_1 \supset G' \\[3mm]  G'\subset
E_2 \subset \s \stackrel{\pi_2} \longleftarrow \tilde\m
\stackrel{\tilde\pi_{G''}} \longrightarrow \s \supset E'' \supset G'', 
\end{array} \nq 
for some embedded graphs $E$, $E_1$, $E_2$ and $E''$. The composition $\m\circ
\tilde\m$ is given by the fibered
product \bq \label{fibLL} \m\circ \tilde\m  := \m\times_{G'} \tilde
\m , \nq with \bq \label{fibLL2} \m\times_{G'} \tilde \m:= \{
(x,y)\in \m\times \tilde \m \,|\, \pi_1(x) =\pi_2(y)
\}. \nq
\end{defn}

First we check that this indeed defines an element $\m\circ \tilde\m \in 
\cC(G,G'')$.

\begin{lem}\label{composeL12}
The composition $\hat \m=\m\times_{G'}\tilde \m$ is a branched cover
$$ E\cup \pi_G\pi_1^{-1}(E_2) \subset S^3
\stackrel{\hat\pi_G}{\leftarrow} \hat \m
\stackrel{\hat\pi_{G''}}{\rightarrow} S^3 \supset E''\cup
\pi_{G''}\pi_2^{-1}(E_1), $$
which defines an element in $\cC(E,E'')$.
If $n$ and $m$ are the generic multiplicities of the covering maps $\pi_G$
and $\pi_1$ of $\m$ and $\tilde n$ and $\tilde m$ are the generic multiplicities for
$\pi_2$ and $\pi_{G''}$, respectively, then the covering maps $\hat\pi_G$
and $\hat\pi_{G''}$ have generic multiplicities $n\tilde n$ and $m \tilde m$.
\end{lem}

\proof Consider the projections $P_1: \m \times_{G'}\tilde \m
\to \m$ and $P_2: \m\times_{G'}\tilde \m \to \tilde \m$. They are
branched covers, respectively branched over $\pi_1^{-1}(E_2)$ and
$\pi_2^{-1}(E_1)$, of order $\tilde n$ and $m$, respectively.
In fact, we have
$$ P_1^{-1}(x)=\{ y\in \tilde M \, |\, 
\pi_2(y)= \pi_1(x)\}. $$ 
Thus, the map $P_1$ is branched over the points $x\in \m$ such that 
$\pi_1(x)$
lies in the branch locus of the map $\pi_2$, that is, the points of 
$\pi_1^{-1}(E_2)\subset \m$. Similarly, the branch locus of the map
$P_2$ is the set $\pi_2^{-1}(E_1) \subset \tilde \m$.
Under the PL covering maps the preimages $\pi_1^{-1}(E_2)$
and $\pi_2^{-1}(E_1)$ are embedded graphs in $\m$ and $\tilde \m$, 
respectively. 

The composite map $\hat\pi_G=\pi_G\circ P_1: \hat \m \to \s$
is branched over the set $E\cup \pi_G\pi_1^{-1}(E_2)$ and the map
$\hat\pi_{G''}=\pi_{G''}\circ P_2: \hat \m \to \s$ is branched over
$E''\cup \pi_{G''}\pi_2^{-1}(E_1)$. Again, because the covering maps 
are PL maps, the sets $\pi_G\pi_1^{-1}(E_2)$ and $\pi_{G''}\pi_2^{-1}(E_1)$ 
are also 1-complexes (graphs) in $\s$ and so are the resulting branch loci. 
Thus, the fibered product $\hat \m$ defines an element of $\cC(G,G'')$.

For the multiplicites at the generic point,
one can just observe that the covering maps $P_1$ and $P_2$ have generic multiplicities
respectively equal to $\tilde n$ and $\tilde m$ so that the composite maps $\pi_G\circ P_1$
and $\pi_{G''}\circ P_2$ have generic multiplicities $n \tilde n$ and $m\tilde m$. 
\endproof

One can similarly derive the formula for the multiplicities over the branch locus and 
the branching indices that count how many branches of the covering come together 
over components of the branch locus.
A simple explicit example of the composition law, in the simplest case of branching
over the unknot, is given by the following.

\begin{exa}\label{cyclexacomp}
Let $\m (n)$ denote the $n$-fold branched cyclic cover of $\s$ branched along the unknot 
$G=O$. The composition $\m (m) \circ \m (n)$ is the cyclic branched
cover $\m (mn)$, viewed as a correspondence in $\cC(O,O)$.
\end{exa}

We then show that the composition is associative. Consider
elements $\m_i\in \cC(G_i,G_{i+1})$, $i=1,2,3$, where we 
use the following notation for the embedded graphs and the branched 
covering maps:
\begin{equation}\label{M123}
\begin{array}{c}
G_1 \subset E_1 \subset S^3 \stackrel{\pi_{11}}{\leftarrow} \m_1
\stackrel{\pi_{12}}{\rightarrow} S^3 \supset E_2 \supset G_2 \\[2mm]
G_2 \subset E_2' \subset  S^3 \stackrel{\pi_{22}}{\leftarrow} \m_2
\stackrel{\pi_{23}}{\rightarrow} S^3 \supset E_3 \supset G_3 \\[2mm]
G_3 \subset E_3' \subset  S^3 \stackrel{\pi_{33}}{\leftarrow} \m_3
\stackrel{\pi_{34}}{\rightarrow} S^3 \supset E_4 \supset G_4.
\end{array}
\end{equation}
We then have the following result.

\begin{prop}\label{assoc}
The composition is associative, namely
\begin{equation}\label{assocM123}
\m_1\circ(\m_2 \circ \m_3) = (\m_1\circ \m_2)\circ \m_3.
\end{equation}
\end{prop}

\proof Consider first the composition $\hat \m_{23}:= \m_2 \circ
\m_3=\m_2\times_{G_2} \m_3$. By Lemma \ref{composeL12}, it is 
a branched cover
$$ \hat E_2 \subset S^3 \stackrel{\hat\pi_{232}}{\leftarrow} \hat
\m_{23} \stackrel{\hat\pi_{234}}{\rightarrow} S^3 \supset \hat E_4,
$$ with branch loci
\begin{equation}\label{hatL2hatL4}
\hat E_2 = E_2' \cup \pi_{22}\pi_{23}^{-1}(E_3') \ \ \text{ and } 
\ \ \hat E_4 = E_4 \cup \pi_{34}\pi_{33}^{-1}(E_3).
\end{equation}

Then the composition $\hat \m_{1(23)}:= \m_1 \circ \hat \m_{23}=
\m_1\circ(\m_2\circ \m_3)$ is a covering
$$
J_1 \subset S^3 \stackrel{\hat\pi_{J_1}}{\leftarrow} \hat \m_{1(23)}
\stackrel{\hat\pi_{J_4}}{\rightarrow} S^3 \supset J_4,
$$
with branch loci 
\begin{equation}\label{linksJ14}
J_1 = E_1 \cup \pi_{11}\pi_{12}^{-1}(\hat E_2) \ \ \ \ J_4= \hat E_4
\cup \hat\pi_{234}\hat\pi_{232}^{-1}(E_2).
\end{equation}

Consider now the composition $\hat \m_{12}:= \m_1\circ \m_2$.
By Lemma \ref{composeL12} above, this is a branched cover
$$ \hat E_1 \subset S^3 \stackrel{\hat\pi_{121}}{\leftarrow} \hat
\m_{12} \stackrel{\hat\pi_{123}}{\rightarrow} S^3 \supset \hat E_3
$$ where $\hat E_1$ and $\hat E_2$ are given by
\begin{equation}\label{hatL1hatL3}
\hat E_1 = E_1 \cup \pi_{11}\pi_{12}^{-1}(E_2') \ \ \ \ \hat E_3 =
E_3 \cup \pi_{23}\pi_{22}^{-1}(E_2).
\end{equation}
Then the composition $\hat \m_{(12)3}:= \hat \m_{12} \circ \m_3=
(\m_1\circ \m_2)\circ \m_3$ is a branched covering
$$
I_1 \subset S^3 \stackrel{\hat\pi_{I_1}}{\leftarrow} \hat \m_{(12)3}
\stackrel{\hat\pi_{I_4}}{\rightarrow} S^3 \supset I_4,
$$
with branch locus 
\begin{equation}\label{linksI14}
I_1 = \hat E_1 \cup \hat\pi_{121}\hat\pi_{123}^{-1}(E_3')  \ \ \ \
I_4= E_4 \cup \pi_{34}\pi_{33}^{-1}( \hat E_3). 
\end{equation}

We have
$$ \hat\pi_{121}\hat\pi_{123}^{-1}(E_3') =
\pi_{11}\pi_{12}^{-1}\pi_{22}\pi_{23}^{-1}(E_3'), $$ so that the
branch loci $J_1=I_1$ agree.
Similarly, we have 
$$ \hat\pi_{234}\hat\pi_{232}^{-1}(E_2) =
\pi_{34}\pi_{33}^{-1}\pi_{23}\pi_{22}^{-1}(E_2) $$ so that the
branch loci $J_4=I_4$ also coincide. A direct computation, using
this same argument, shows that the multiplicities of the covering 
maps also agree, as well as the branching indices. 
Thus, the manifolds $\m_1\circ(\m_2 \circ \m_3)$ 
and $(\m_1\circ \m_2)\circ \m_3$ are the same as branched covers. 
\endproof

\medskip
\subsection{The unit of composition}

Let $\bU$ denote the trivial unbranched covering $id: \s \to \s$, 
viewed as an element $\bU_G \in \cC(G,G)$ for any embedded graph $G$. 
We have the following result.

\begin{lem}\label{idmorph}
The trivial covering $\bU$ is the identity element
for composition.
\end{lem} 

\proof Consider the composition $\m\circ \bU_{G'}$. The fibered
product satisfies $$\m\times_{G'} \s = \{(m,s)\in \m \times \s \,|\,
\pi_2(m)=s \} =\bigcup_{s\in \s}\pi_2^{-1}(s)=\m.$$ So the
projection map $P_1 :\m\times_{G'} \s \to \m$ is just the identity map 
$id: \m \to \m$, with
the composite map $\hat{\pi}_G=\pi_1\circ P_1=\pi_1$. The projection
map $P_2: \m\times_{G'} \s\to \s$, sending $(m,s)\mapsto s$ for
$m\in \pi_2^{-1}(s)$, is just the map $P_2=\pi_2$, hence
$\hat{\pi}_G=\pi_4\circ P_2=\pi_2$. Thus, we see that $\m\times_G
\s=\m$ with $\pi_G=\pi_1$ and $\pi_{G'}=\pi_2$. This shows that
$\m\circ \bU_{G'}=\m$. 
The argument for the composition $\bU_G \circ \m$ is analogous.
\endproof

\section{Semigroupoids and additive categories}\label{CatSec}

A semigroupoid (\cf \cite{Hu}) is a collection $\cG$ with a 
partially defined associative product. An element $\gamma \in \cG$ is a
unit if $\gamma \alpha =\alpha$ and $\beta\gamma =\beta$
for all $\alpha$ and $\beta$ in $\cG$ for which the product 
is defined. We denote by $\cU(\cG)$ the set of units of $\cG$.
A semigroupoid is regular if, for all $\alpha\in\cG$
there exist units $\gamma$ and $\gamma'$ such that $\gamma \alpha$ 
and $\alpha \gamma'$ are defined. Such units, if they exist, are unique.
We denote them by $s(\alpha)$ (the source) and $r(\alpha)$ (the range).
They satisfy $s(\alpha\beta)=s(\alpha)$ and $r(\alpha\beta)=r(\beta)$.
To each unit $\gamma \in \cU(\cG)$ in a regular semigroupoid one 
associates a subsemigroupoid
$\cG_\gamma =\{ \alpha \in \cG \,|\, r(\alpha)=\gamma \}$.

\smallskip
  
We can reformulate the results on embedded graphs and 3-manifolds obtained 
in the previous section in terms of semigroupoids in the following way. 

\begin{lem}\label{3mfldG}
The set of closed oriented 3-manifolds forms a regular
semigroupoid, whose set of units is identified with the set of 
embedded graphs.
\end{lem}

\proof
We let $\cG$ be the collection of data $\alpha=(\m,G,G')$ with $\m$ a closed 
oriented 3-manifold with branched covering maps to $\s$
of the form \eqref{2covers}. We define a composition rule as in Definition
\ref{comp2}, given by the fibered product. In the multi-connected case, for 
\begin{equation}\label{Mmultconn}
M=M_1\amalg M_2\amalg \cdots \amalg M_k
\end{equation}
with $(M_i,G,G')$ as in \eqref{2covers} with $M_i$ connected, we extend 
the composition $M\circ \tilde M$ to mean
\begin{equation}\label{circMmultconn}
M\circ \tilde M =M_1\circ \tilde M\amalg M_2\circ \tilde M\amalg \cdots \amalg M_k\circ \tilde M, 
\end{equation}
and similarly for $\tilde M$ multi-connected. It is necessary to include the
multi-connected case since the fibered product of connected manifolds
may consist of different connected components. We impose the condition that the 
composition of $\alpha=(\m_1,G_1,G'_1)$ and $\beta=(\m_2,G_2,G_2')$
is only defined when the $G_1'=G_2$.

By Lemma \ref{idmorph}, we know that, for each $\alpha=(\m,G,G')\in \cG$
the source and range are given by the trivial coverings 
$\gamma=\bU_G=(\bU,G,G)$ and $\gamma'=\bU_{G'}=(\bU,G',G')$. That is, we can 
identify them with $s(\alpha)=G$ and $r(\alpha)=G'$. Thus, the set of units 
$\cU(\cG)$ is the set of embedded graphs in $\s$.
\endproof

For a given embedded graph $G$, the subsemigroupoid $\cG_G$ is given by
the set of all 3-manifolds that are covering of $\s$ branched along
embedded graphs $E$ containing $G$ as a subgraph. 

\smallskip

Given a semigroupoid $\cG$, and a commutative ring $R$, one can define 
an associated semigroupoid ring $R[\cG]$, whose elements are finitely
supported functions $f: \cG \to R$, with the associative product
\begin{equation}\label{semigrprod}
(f_1 * f_2)(\alpha) = \sum_{\alpha_1,\alpha_2 \in \cG : 
\alpha_1\alpha_2=\alpha} f_1(\alpha_1) f_2(\alpha_2) .
\end{equation}

Elements of $R[\cG]$ can be equivalently described as finite $R$-combinations 
of elements in $\cG$, namely $f=\sum_{\alpha \in \cG} a_\alpha \delta_\alpha$, 
where $a_\alpha =0$ for all but finitely many $\alpha\in \cG$ and 
$\delta_\alpha(\beta)=\delta_{\alpha,\beta}$, the Kronecker delta.

In the multi-connected case $\alpha=(\m,G,G')$ with $M$ as in \eqref{Mmultconn} 
we impose the relation 
\begin{equation}\label{deltarels}
\delta_\alpha = \sum_{i=1}^k \delta_{\alpha_i},
\end{equation}
where $\alpha_i=(M_i,G,G')$ with $M_i$ connected.

\smallskip

The following statement is a semigroupoid version 
of the representations of groupoid algebras generalizing
the regular representation of group rings.

\begin{lem}\label{repVgamma}
Suppose given a unit $\gamma \in \cU(\cG)$. Let $\cH_\gamma$ denote the
$R$-module of finitely supported functions $\xi : \cG_\gamma \to R$.
The action
\begin{equation}\label{actVgamma}
\rho_\gamma(f)(\xi)(\alpha)=\sum_{\alpha_1\in \cG, \alpha_2\in \cG_\gamma: 
\alpha=\alpha_1 \alpha_2} f(\alpha_1) \xi(\alpha_2),
\end{equation}
for $f\in R[\cG]$ and $\xi \in \cH_\gamma$,
defines a representation of $R[\cG]$ on $\cH_\gamma$.
\end{lem}

\proof We have 
$$ \rho_\gamma(f_1*f_2)(\xi)(\alpha)=\sum (f_1*f_2)(\alpha_1)\xi(\alpha_2) $$
$$ = \sum_{\beta_1\beta_2=\alpha_1 \in \cG} \sum_{\alpha_1 \alpha_2=\alpha} 
f_1(\beta_1) f_2(\beta_2) \xi(\alpha_2) = \sum_{\beta_1\beta=\alpha}  
f_1(\beta_1) \rho_\gamma(f_2)(\xi)(\beta), $$
hence $\rho_\gamma(f_1*f_2)=\rho_\gamma(f_1)\rho_\gamma(f_2)$.
Since for elements of a semi-groupoid the range satisfies $r(\alpha\beta)=r(\beta)$,
the action is well defined on $\cH_\gamma$.
\endproof

In the next section we see that the fact the difference in the representation
\eqref{actVgamma} between the semigroupoid and the groupoid case manifests 
itself in the compatibility with the involutive structure. 

\medskip

A semigroupoid is just an equivalent formulation of a small category, 
so the result above simply states that embedded graphs form a small 
category with the sets $\cC(G,G')$ as morphisms. Passing from the 
semigroupoid $\cG$ to $R[\cG]$ corresponds to passing from a small 
category to its additive envelope, as follows.

\smallskip

Let $R$ be a commutative ring. We replace the sets $\cC(G,G')$ of 
geometric correspondences by $R$-modules.

\begin{defn}\label{HomKKprime}
For given embedded graphs $G$ and $G'$ in $\s$, let $\Hom_R(G,G')$ denote
the free $R$-module generated by the elements of $\cC(G,G')$. 
\end{defn}

Namely, elements $\phi \in \Hom_R(G,G')$ are finite $R$-combinations
$\phi=\sum_\m a_\m \m$, where the sum ranges over the set of all 
3-manifolds that are branched covers as in \eqref{2covers} and the
$a_\m \in R$ satisfy $a_\m=0$ for all but finitely many $\m$. 
Again, in the case of geometric correspondences given by multi-connected 
manifolds, we impose the relation $\m=\m_1 +\m_2$ when
$\m = \m_1 \amalg \m_2$.

For $R=\Z$, we simply write $\Hom(G,G')$ for $\Hom_\Z(G,G')$. It then follows 
immediately that we obtain in this way a pre-additive category.

\begin{lem}\label{preadd}
The category $\cK$ whose objects are embedded graphs and with morphisms 
the $\Hom(G,G')$ is a pre-additive category.
\end{lem}

One can pass to its additive closure by considering the category
$Mat(\cK)$ whose objects are formal direct sums of objects of $\cK$
and whose morphisms are matrices of morphisms in $\cK$.
In the following we continue to use the notation $\cK$ for the 
additive closure. For $R=k$ a field, we obtain in this way a 
$k$-linear category, where morphism spaces are $k$-vector spaces.

\section{Convolution algebra and time evolutions}\label{timeSect}

Consider as above the semigroupoid ring $\C[\cG]$ of complex valued functions with
finite support on $\cG$, with the associative convolution product \eqref{semigrprod},
\begin{equation}\label{semigrprod2}
(f_1 * f_2)(\m) = \sum_{\m_1,\m_2 \in \cG : 
\m_1\circ\m_2=\m} f_1(\m_1) f_2(\m_2) .
\end{equation}

We define an involution on the semigroupoid $\cG$ by setting
\begin{equation}\label{alphavee}
\cC(G,G') \ni \alpha=(\m,G,G') \mapsto \alpha^\vee=(\m,G',G) \in \cC(G',G),
\end{equation}
where, if $\alpha$ correposnds to the 3-manifold $\m$ with branched covering maps
$$ G\subset E \subset \s \stackrel{\pi_G}{\leftarrow} \m
\stackrel{\pi_{G'}}{\rightarrow} \s \supset E' \supset G' $$
then $\alpha^\vee$ corresponds to the same 3-manifold with maps
$$ G'\subset E' \subset \s \stackrel{\pi_{G'}}{\leftarrow} \m
\stackrel{\pi_G}{\rightarrow} \s \supset E \supset G $$
taken in the opposite order.
In the following, for simplicity of notation, we write $\m^\vee$ 
instead of $\alpha^\vee=(\m,G',G)$. 

\begin{lem}\label{invol1}
The algebra $\C[\cG]$ is an involutive algebra with the involution
\begin{equation}\label{inv1}
f^\vee(\m)=\overline{f(\m^\vee)}.
\end{equation}
\end{lem}

\proof We clearly have $(af_1 + bf_2)^\vee = \bar a f_1^\vee + \bar b
f_2^\vee$ and $(f^\vee)^\vee=f$. We also have
$$  (f_1 * f_2)^\vee (\m)= \sum_{\m^\vee =\m_1^\vee \circ \m_2^\vee}
\overline{f_1}(\m_1^\vee) \overline{f_2}(\m_2^\vee) =
\sum_{\m = \m_2\circ \m_1}  f_2^\vee (\m_2) f_1^\vee(\m_1) $$
so that $(f_1 * f_2)^\vee = f_2^\vee * f_1^\vee$.
\endproof

\subsection{Time evolutions}

Given an algebra $\cA$ over $\C$, a time evolution is a 1-parameter
family of automorphisms 
$\sigma: \R \to \Aut(\cA)$.  There is a natural time evolution on the
algebra $\C[\cG]$  
obtained as follows.

\begin{lem}\label{timeevlem}
Suppose given a function $f\in \C[\cG]$. Consider the action defined by
\begin{equation}\label{sigmatnm}
\sigma_t(f)(\m) := \left(\frac{n}{m}\right)^{it} f(\m),
\end{equation}
where $\m$ a covering as in \eqref{2covers}, with the covering maps
$\pi_G$ and $\pi_{G'}$ respectively of generic multiplicity $n$ and $m$. 
This defines a time evolution on $\C[\cG]$. 
\end{lem}

\proof Clearly $\sigma_{t+s}=\sigma_t\circ\sigma_s$. We check that
$\sigma_t(f_1*f_2)=\sigma_t(f_1)*\sigma_t(f_2)$. By
\eqref{semigrprod2}, we have 
$$ \sigma_t(f_1*f_2)(\m)= \left(\frac{n}{m}\right)^{it} (f_1 * f_2)(\m) $$
$$ = \sum_{\m_1,\m_2 \in \cG : 
\m_1\circ\m_2=\m}  \left(\frac{n_1}{m_1}\right)^{it} f_1(\m_1)
\left(\frac{n_2}{m_2}\right)^{it} 
 f_2(\m_2) =(\sigma_t(f_1)*\sigma_t(f_2))(\m), $$
where $n_i,m_i$ are the generic multiplicities of the covering maps
for $\m_i$, with $i=1,2$. In fact, 
we know by Lemma \ref{composeL12} that $n=n_1n_2$ and $m=m_1 m_2$.
The time evolution is compatible with the involution \eqref{inv1}, since we have 
$$ \sigma_t(f^\vee)(\m)=\left(\frac{n}{m}\right)^{it} f^\vee(\m)= \left(\frac{n}{m}\right)^{it}
\overline{f(\m^\vee)}=\overline{\left(\frac{m}{n}\right)^{it}  f(\m^\vee)}=
\overline{\sigma_t(f)(\m^\vee)}=(\sigma_t(f))^\vee(\m). $$
\endproof

Similarly, we define the left and right time evolutions on $\cA$ by setting
\begin{equation}\label{LRtime}
\sigma^L_t(f)(M):= n^{it} \, f(M), \ \ \ \ \  \sigma^R_t(f)(M):= m^{it}\, f(M),
\end{equation}
where $n$ and $m$ are the multiplicities of the two covering maps as above.
The same argument of Lemma \ref{timeevlem} shows that the $\sigma^{L,R}_t$ are
time evolutions. One sees by construction that they commute, \ie that  
$[\sigma_t^L,\sigma_t^R]=0$.
The time evolution \eqref{sigmatnm} is the composite
\begin{equation}\label{sigmaLRprod}
\sigma_t = \sigma^L_t \sigma^R_{-t}.
\end{equation}
The involution exchanges the two time evolutions by
\begin{equation}\label{involLR}
\sigma^L_t(f^\vee)=(\sigma^R_{-t} (f))^\vee .
\end{equation}

\subsection{Creation and annihilation operators}

Given an embedded graph $G\subset \s$, consider, as above, the set $\cG_G$ of all
3-manifolds that are branched covers of $\s$ branched along an embedded graph 
$E\supset G$.  

On the vector space $\cH_G$ of finitely supported complex valued functions 
on $\cG_G$ we have a representation of $\C[\cG]$ as in Lemma \ref{repVgamma}, 
defined by 
\begin{equation}\label{repCG}
(\rho_G(f) \xi) (\m) = \sum_{\m_1\in \cG,\m_2 \in \cG_G : 
\m_1\circ \m_2=\m} f(\m_1) \xi(\m_2).
\end{equation}

It is natural to consider on the space $\cH_G$ the inner product
\begin{equation}\label{innprodHG}
\langle \xi, \xi'\rangle = \sum_{\m\in \cG_G} \overline{\xi(\m)} \xi'(\m).
\end{equation}

Notice however that, unlike the usual case of groupoids, the involution 
\eqref{inv1} given by the transposition of the correspondence does not 
agree with the adjoint in the inner product \eqref{innprodHG}, namely
$\rho_\gamma(f)^* \neq \rho_\gamma(f^\vee)$. 

The reason behind this incompatibility is that semigroupoids behave 
like semigroup algebras implemented by isometries rather 
than like group algebras implemented by unitaries. The model case for an adjoint 
and involutive structure that is compatible with the representation \eqref{repCG} 
and the pairing \eqref{innprodHG} is therefore given by the algebra of creation
and annihilation operators. 

We need the following preliminary result.

\begin{lem}\label{division}
Suppose given elements $\alpha=(\m,G,G')$ and $\alpha_1=(\m_1,G_1,G_1')$ in $\cG$. 
If there exists an element $\alpha_2=(\m_2,G_2,G_2')$ in $\cG(G_2,G_2')$ such
that $\alpha=\alpha_1\circ\alpha_2 \in \cG$, then $\alpha_2$ is unique.
\end{lem}

\proof We have $\m=\m_1\circ \m_2$. We denote by $E\supset G$, $E'\supset G'$ and
$E_1\supset G_1$ and $E_1'\supset G_1'$ the embedded graphs that are the branching
loci of the covering maps $\pi_G$, $\pi_{G'}$ and $\pi_{G_1}$, $\pi_{G_1'}$ of $\m$ 
and $\m_1$, respectively.

By construction we know that for the composition $\alpha_1\circ\alpha_2$ to be
defined in $\cG$ we need to have $G_1'=G_2$. Moreover, by
Lemma \ref{composeL12} we know that $E=E_1 \cup \pi_{G_1} \pi_{G_1'}^{-1}(E_2)$
and $E'= E_2' \cup \pi_{G_2'}\pi_{G_2}^{-1}(E_1')$, where $E_2$ and $E_2'$ are the
branch loci of the two covering maps of $\m_2$. 

The manifold $\m_2$ and the branched covering maps $\pi_{G_2}$ and $\pi_{G_2'}$
can be reconstructed by determining the multiplicities, branch indices, and 
branch loci $E_2$, $E_2'$. 

The $n$-fold branched covering $\pi_G: \m \to S^3 \supset E\supset G$ 
is equivalently described by a representation of the fundamental group 
$\pi_1(S^3 \smallsetminus E) \to S_n$.
Similarly, the $n_1$-fold branched covering $\pi_{G_1}: \m_1 \to S^3 \supset E_1
\supset G_1$ is specified by a representation $\pi_1(S^3\smallsetminus E_1) \to S_{n_1}$. 
Given these data, we obtain the branched covering $P_1: \m \to \m_1$ such that
$\pi_G=\pi_{G_1}\circ P_1$ in the following way. The restrictions
$\pi_G: \m\smallsetminus \pi_G^{-1}(E) \to S^3 \smallsetminus E$ and
$\pi_{G_1}: \m_1\smallsetminus \pi_{G_1}^{-1}(E) \to S^3\smallsetminus E$ are
ordinary coverings, and we obtain from these the covering 
$P_1: \m\smallsetminus \pi_G^{-1}(E) \to  \m_1\smallsetminus \pi_{G_1}^{-1}(E)$.
Since this is defined on the complement of a set of codimension two, it extends uniquely 
to a branched covering $P_1: \m\to \m_1$.
The image under $\pi_{G_1'}$ of the branch locus of $P_1$ and the multiplicities and 
branch indices of $P_1$ then determine uniquely the manifold $\m_2$ as a branched covering
$\pi_{G_2}: \m_2 \to S^3 \supset E_2$. Having determined the branched covering
$\pi_{G_2}$ we have the covering maps realizing $\m$ as the fibered product of $\m_1$ 
and $\m_2$, hence we also have the branched covering map $P_2: \m \to \m_2$. 

The knowledge of the branch loci, multiplicities and branch indices
of $\pi_{G'}$ and $P_2$ then allows us to identify the part of the
branch locus $E'$ that constitutes $E_2'$ and the multiplicities
and branch indices of the map $\pi_{G_2'}$. This completely
determines also the second covering map $\pi_{G_2'}: \m_2 \to S^3 \supset E_2'$.
\endproof

We denote in the following by the same notation $\cH_G$ the Hilbert space completion 
of the vector space $\cH_G$ of finitely supported complex valued functions 
on $\cG_G$ in the inner product \eqref{innprodHG}. We denote by $\delta_\m$ the
standard orthonormal basis consisting of functions $\delta_\m(\m')=\delta_{\m,\m'}$,
with $\delta_{\m,\m'}$ the Kronecker delta.

\smallskip

Given an element $\m\in \cG$, we define an associated bounded linear
operator $A_\m$ on $\cH_G$ of the form 
\begin{equation}\label{AMop}
(A_\m \xi)(\m') = \left\{ \begin{array}{ll} \xi(\m'') & \text{if } \, \m'=\m\circ\m'' 
\\[2mm] 0 & \text{otherwise.} \end{array}\right.
\end{equation}
Notice that \eqref{AMop} is well defined because of Lemma \ref{division}. 

\smallskip

\begin{lem}\label{AMstarLem}
The adjoint of the operator \eqref{AMop} in the inner product \eqref{innprodHG} 
is given by the operator
\begin{equation}\label{AMstar}
(A_\m^* \xi)(\m') = \left\{ \begin{array}{ll} \xi(\m \circ \m') & 
\text{if the composition is defined} 
\\[2mm] 0 & \text{otherwise.} \end{array}\right.
\end{equation}
\end{lem}

\proof We have
$$ \langle \xi, A_\m \zeta \rangle =\sum_{\m'=\m\circ \m''} 
\overline{\xi(\m')} \zeta(\m'') =\sum_{\m''} \overline{\xi(\m\circ \m'')} \zeta(\m'')
=\langle A_\m^* \xi, \zeta \rangle. $$
\endproof

We regard the operators $A_\m$ and $A_\m^*$ as the annihilation and creation operators
on $\cH_G$ associated to the manifold $\m$. They satisfy the following relations.

\begin{lem}\label{anncreops}
The products $A_\m^* A_\m =P_\m$ and $A_\m A_\m^*=Q_\m$ are given, respectively, by the 
projection $P_\m$ onto the subspace of $\cH_G$ given by the range of 
composition by $\m$, and the projection $Q_\m$ onto the subspace of $\cH_G$ spanned by
the $\m'$ with $s(\m')=r(\m)$. 
\end{lem}

\proof This follows directly from \eqref{AMop} and \eqref{AMstar}.
\endproof

The following result shows the relation between the algebra $\C[\cG]$ and the
algebra of creation and annihilation operators $A_\m$, $A_\m^*$.

\begin{lem}\label{CGandAM}
The algebra of linear operators on $\cH_G$ generated by the $A_\m$ is the image 
$\rho_G(\C[\cG])$ of $\C[\cG]$ under the representation $\rho_G$ of \eqref{repCG}.
\end{lem}

\proof Every function $f\in \C[\cG]$ is by construction a finite linear combination
$f=\sum_\m a_\m \delta_\m$, with $a_\m\in R$. Under the representation $\rho_G$ we have
\begin{equation}\label{rhoAM}
 (\rho_G(\delta_\m)\xi)(\m')=\sum_{\m'=\m_1 \circ \m_2} \delta_\m(\m_1) \xi(\m_2)=
(A_\m \xi)(\m'). 
\end{equation}
\endproof

This shows that, when working with the representations $\rho_G$ the correct way to obtain
an involutive structure is by extending the algebra generated 
by the $A_\m$ to include the $A_\m^*$, instead of using the involution \eqref{inv1}
of $\C[\cG]$. 

\subsection{Time evolutions and Hamiltonians}

Given a representation $\rho : \cA \to \End(\cH)$ of an algebra $\cA$
with a time evolution $\sigma$, 
one says that the time evolution, in the representation $\rho$, is
generated by a Hamiltonian $H$ if 
for all $t\in \R$ one has
\begin{equation}\label{Ham}
\rho(\sigma_t(f))=e^{-itH} \rho(f) e^{itH},
\end{equation}
for an operator $H \in \End(\cH)$.  

\begin{lem}\label{timextend}
The time evolutions $\sigma^L_t$ and $\sigma^R_t$ of \eqref{LRtime} and $\sigma_t=\sigma^L_t
\sigma^R_{-t}$ of \eqref{sigmatnm} extend to time evolutions of the involutive algebra generated by 
the operators $A_\m$ and $A_\m^*$ by
\begin{equation}\label{AMsigma}
\begin{array}{ll}
\sigma_t^L(A_M)=n^{it} A_M & \sigma_t^L(A_M^*)=n^{-it} A_M^* \\[2mm]
\sigma_t^R(A_M)=m^{it} A_M & \sigma_t^R(A_M^*)=m^{-it} A_M^* \\[2mm]
\sigma_t(A_M)=\left(\frac{n}{m}\right)^{it} A_M & \sigma_t(A_M^*)=\left(\frac{n}{m}\right)^{-it} A_M^*.
\end{array}
\end{equation}
\end{lem}

\proof The result follows directly from \eqref{rhoAM} and the condition 
$\sigma_t(T^*)=(\sigma_t(T))^*$.
\endproof

We then have immediately the following result. We state it for the time evolution
$\sigma_t^L$, while the case of $\sigma^R_t$ is analogous.

\begin{lem}\label{HamM}
Consider the unbounded linear operators $H^L_{G'}$ and $H^R_{G'}$ on the space 
$\cH_{G'}$ defined by
\begin{equation}\label{Hop}
(H^L_{G'}\, \xi)(\m) = \log(n) \,\, \xi(\m), \ \ \ \ \ (H^R_{G'}\, \xi)(\m) = \log(m) \,\, \xi(\m)
\end{equation}
for $\m$ a geometric correspondence of the form
$$ G \subset E \subset \s \stackrel{\pi_G}{\longleftarrow} \m
\stackrel{\pi_{G'}}{\longrightarrow}  \s \supset E' \supset G'
$$
with $\pi_G$ and $\pi_{G'}$ branched coverings of order $n$ and $m$, respectively. 
Then $H^L_{G'}$ and $H^R_{G'}$ are, respectively, Hamiltonians for the time evolutions 
$\sigma^L_t$ and $\sigma_t^R$ in the representation $\rho_{G'}$ of \eqref{repCG}.
\end{lem}

\proof It is immediate to check that $\rho_{G'}(\sigma^L_t(f))=e^{-itH^L} \rho_{G'}(f) e^{itH^L}$ and
$\rho_{G'}(\sigma^R_t(f))=e^{-itH^R} \rho_{G'}(f) e^{itH^R}$
for $f\in \C[\cG]$. In fact, it suffices to use the explicit form of the time evolutions on the 
creation and annihilation operators given in Lemma \ref{timextend} above to see that
they are implemented by the Hamiltonians $H_{G'}^L$ and $H_{G'}^R$.
\endproof

An obvious problem with these time evolutions is the fact that the corresponding
Hamiltonians typically can have infinite multiplicities of the
eigenvalues. For example, by the strong form of the Hinden-Montesinos
theorem \cite{pr} and the existence of universal knots \cite{HLM},
there exist knots $K$ such that all closed oriented 3-manifolds can
be obtained as a 3-fold branched cover of $\s$, branched along
$K$. For this reason it is useful to consider time
evolutions on a convolution algebra of geometric correspondences that
takes into account the equivalence given by 4-dimensional
cobordisms. We turn to this in \S \ref{CobordSec} and \S
\ref{2catSec} below. 

\section{Cobordisms and equivalence of correspondences}\label{CobordSec}

Whenever one defines morphisms via
correspondence, be it cycles in the product as in the case of
motives or submersions as in the case of geometric correspondences of \cite{CoSka},
the most delicate step is always deciding up to what equivalence
relation correspondences should be considered. 
In the case of 3-manifolds with the structure of branched covers,
there is a natural notion of equivalence, which is given by
cobordisms of branched covers, \cf \cite{HL}. Adapted to our setting,
this is formulated in the following way.

\begin{defn}\label{cobord}
Suppose given two correspondences $\m_1$ and $\m_2$ in $\cC(G,G')$, of
the form
$$ G \subset E_1 \subset \s \stackrel{\pi_{G,1}}{\longleftarrow} \m_1
\stackrel{\pi_{G',1}}{\longrightarrow}  \s \supset E'_1 \supset G'
$$
$$ G \subset E_2 \subset \s \stackrel{\pi_{G,2}}{\longleftarrow} \m_2
\stackrel{\pi_{G',2}}{\longrightarrow}  \s \supset E'_2 \supset G'.
$$ 
Then a cobordism between $\m_1$ and $\m_2$ is a $4$-dimensional PL
manifold $W$ with boundary $\partial W= \m_1 \cup -\m_2$, endowed
with two branched covering maps
\begin{equation}\label{Wmaps}
S \subset \s\times [0,1] \stackrel{q}{\longleftarrow} W
\stackrel{q'}{\longrightarrow} \s\times [0,1] \supset S',
\end{equation}
branched along surfaces (PL embedded 2-complexes) $S,S'\subset \s\times [0,1]$. 
The maps $q$ and $q'$ have the properties that $\m_1 = q^{-1}(\s\times \{0\})=
q'^{-1}(\s\times \{0\})$ and $\m_2 =q^{-1}(\s\times \{1\})=
q'^{-1}(\s\times \{1\})$, with $q|_{\m_1}=\pi_{G,1}$,
$q'|_{\m_1}=\pi_{G',1}$, $q|_{\m_2}=\pi_{G,2}$ and
$q'|_{\m_2}=\pi_{G',2}$. The surfaces $S$ and $S'$ have boundary
$\partial S= E_1\cup -E_2$ and $\partial S'=E_1'\cup -E_2'$, with
$E_1= S \cap (\s\times \{0\})$, $E_2=S \cap (\s\times \{1\})$,
$E_1'=S' \cap (\s\times \{0\})$, and $E_2'=S' \cap (\s\times
\{1\})$.
\end{defn}

\begin{lem}\label{equivrel}
Setting $\m_1 \sim \m_2$ if there exists a cobordims $W$ as in
Definition \ref{cobord} defines an equivalence relation. Moreover, 
if $\m_1 \sim \m_2$ in $\cC(G,G')$ and $\m_1'\sim \m_2'$ in 
$\cC(G',G'')$, then $\m_1 \circ \m_1' \sim \m_2 \circ \m_2'$.
\end{lem}

\proof We have $\m\sim \m$ through the trival cobordism $\m\times [0,1]$.
The symmetric property is satisfied by taking the opposite orientation
cobordims and transitivity is achieved by gluing cobordisms along their
common boundary, $W=W_1\cup_{\m_2}W_2$. This can be done compatibly
with the branched covering maps, since these match along the common
boundary. Thus, we have a well defined
equivalence relation. To check the compatibility with composition, let
\begin{equation}\label{Wmaps1}
S_{11} \subset \s\times [0,1] \stackrel{q_1}{\longleftarrow} W_1
\stackrel{q'_1}{\longrightarrow} \s\times [0,1] \supset S_{12}
\end{equation}
be a cobordism realizing the equivalence $\m_1 \sim \m_2$, and 
\begin{equation}\label{Wmaps133}
S_{21} \subset \s\times [0,1] \stackrel{q_2}{\longleftarrow} W_2
\stackrel{q'_2}{\longrightarrow} \s\times [0,1] \supset S_{22},
\end{equation}
be a cobordims realizing $\m_1' \sim \m_2'$. We check that the
fibered product
\begin{equation}\label{W1circW2}
W_1 \circ W_2 := \{(x,y)\in W_1 \times W_2 | q_1'(x)=q_2(y) \}
\end{equation}
gives a branched covers cobordism realizing the desired
equivalence $\m_1\circ\m'_1 \sim \m_2\circ \m'_2$. First notice that 
we have
$$ \partial(W_1\circ W_2) = \partial W_1 \circ \partial W_2 
= (\m_1 \circ \m'_1 )\cup-(\m_2 \circ \m'_2).
$$
Moreover, $W_1 \circ W_2$ is a branched cover
$$\hat{S}_1 \subset \s \times [0,1]
\stackrel{T_1}{\leftarrow} W_1 \circ W_2 \stackrel{T_2}{\rightarrow}
\s \times [0,1] \supset \hat{S}_2 ,$$
with branch loci $\hat{S}_1=S_{11}\cup q_1(q'^{-1}_1(S_{21}))$ and 
$\hat{S}_2= S_{22}\cup q'_2(q^{-1}_1(S_{12}))$, satisfying
$$ \partial \hat{S_1}=\partial ( S_{11} \cup q_1(q'^{-1}_1(S_{21})))=
E_{11}\cup\pi_{11}(\pi^{-1}_{12}(E'_{11}))\cup (-E_{21}\cup
\pi_{21}(\pi^{-1}_{22}(E'_{21}))) = I_1 \cup -I_3, $$
with the notation of Lemma \ref{composeL12}. Similarly, we 
have $\partial\hat{S_2}=I_2 \cup -I_4$. 
The sets $q_1(q'^{-1}_1(S_{21}))$, $q'_2(q^{-1}_1(S_{12}))$ are
2-complexes in $\s\times [0,1]$ with boundary the 1-complexes
$\pi_{11}(\pi^{-1}_{12}(E'_{11}))$ and
$\pi_{21}(\pi^{-1}_{22}(E'_{21}))$. 
We then see that
$$ T_1^{-1}(\s \times \{0 \})= \m_1 \circ
\m'_1 = T_2^{-1}(\s \times \{0\}) $$
$$ T_1^{-1}(\s \times \{1\})= \m_2 \circ
\m'_2 = T_2^{-1}(\s \times \{1\}). $$
In fact, the first set is equal to 
$$ \begin{array}{l}
\{(x,y)\in  q^{-1}_1(\s\times \{0\})\times W_2 : q'_1(x)= q_2(y)\} = \\
 \{(x,y) \in  q^{-1}_1(\s\times \{0\})\times
q^{-1}_2(\s \times \{0\}): q'_1(x)=q_2(y)\}. \end{array} $$
The other case is analogous.
Thus, the resulting $W_1 \circ W_2$ is a branched cover
cobordism with the desired properties.
\endproof

\medskip

We can now consider the sets of geometric correspondences, up to
the equivalence relation of cobordism. The result of Lemma \ref{equivrel} 
above immediately implies the following.

\begin{lem}\label{CKKmod}
Let $G$ and $G'$ be embedded graphs in $\s$ and let $\cC(G,G')$ be the set of
geometric correspondences as in Definition \ref{CKKprime}. Let
\begin{equation}\label{defCKKmod}
 \cC_\sim (G,G'):= \cC(G,G')/\sim
\end{equation}
denote the quotient of $\cC(G,G')$ by the equivalence relation of
cobordism of Definition \ref{cobord}. There is an induced associative
composition
\begin{equation}\label{compmod}
\circ: \cC_\sim (G,G')\times \cC_\sim (G',G'') \to \cC_\sim (G,G'').
\end{equation} 
\end{lem}

As in \S \ref{CatSec} above, given a commutative ring $R$ we define
$\Hom_{R,\sim}(G,G')$ to be the free $R$-module generated by 
$\cC_\sim (G,G')$, that is, the set of finite $R$-combinations
$\phi = \sum_{[\m]} a_{[\m]} [\m]$, with $[\m] \in \cC_\sim (G,G')$ and
$a_{[\m]}\in R$ with $a_{[\m]}=0$ for all but finitely many $[\m]$.
We write $\Hom_\sim(G,G')$ for $\Hom_{\Z,\sim}(G,G')$.

We then construct a category $\cK_{R,\sim}$ of embedded graphs and correspondences in 
the following way.

\begin{defn}\label{knotmotives}
The category $\cK_{R,\sim}$ has objects the embedded graphs $G$ in $\s$  
and morphisms the $\Hom_{R,\sim}(G,G')$
\end{defn}

After passing to $Mat(\cK_{R,\sim})$ one obtains an additive 
category of embedded graphs and correspondences, which one still denotes 
$\cK_{R,\sim}$. 

\subsection{Time evolutions and equivalence}\label{tevolSubsec}

We return now to the time evolutions \eqref{LRtime} and \eqref{sigmatnm} 
on the convolution algebra $\C[\cG]$.
After passing to equivalence classes by the relation of cobordism, we
can consider the semigroupoid  
$\bar\cG$ which is given by the data $\alpha=([\m],G,G')$, where
$[\m]$ denotes the equivalence 
class of $\m$ under the equivalence relation of branched cover
cobordism. Lemma \ref{equivrel} shows  that the composition in the
semigroupoid $\cG$ induces a well defined composition law in
$\bar\cG$. 
We can then consider the algebra $\C[\bar\cG]$ with the convolution product as in 
\eqref{semigrprod2},
\begin{equation}\label{semigrprod3}
(f_1 * f_2)([\m]) = \sum_{[\m_1],[\m_2] \in \bar\cG : 
[\m_1]\circ[\m_2]=[\m]} f_1([\m_1]) f_2([\m_2]) .
\end{equation}

The involution $f\mapsto f^\vee$ is also compatible with the
equivalence relation, as it extends to the involution on the
cobordisms $W$ that interchanges the two branched covering maps.

\smallskip

\begin{lem}\label{timeevCGbar}
The time evolutions \eqref{LRtime} and \eqref{sigmatnm} descend to well defined time
evolutions on the algebra $\C[\bar\cG]$.
\end{lem}

\proof The result follows from the fact that the generic multiplicity
of a branched covering is invariant under branched cover
cobordisms. Thus, we have an induced time evolution of the form
\begin{equation}\label{sigmatnmbar}
\sigma^L_t(f)[\m] := n^{it} f[\m], \ \ \ \  \sigma^R_t(f)[\m] := m^{it} f[\m], \ \ \ \ 
\sigma_t(f)[\m] := \left(\frac{n}{m}\right)^{it} f[\m],
\end{equation}
where each representative in the class $[\m]$ has branched covering
maps with multiplicities
$$ G \subset E \subset \s \stackrel{n:1}\leftarrow \m
\stackrel{m:1}\rightarrow \s \supset E' \supset G'. $$
We see that the time evolution is compatible with the involution 
as in Lemma \ref{timeevlem}.
\endproof

\subsection{Representations and Hamiltonian}

Similarly, we can again consider representations of $\C[\bar\cG]$ as in \eqref{repCG}
\begin{equation}\label{repCGbar}
(\rho(f) \xi) [\m] = \sum_{[\m_1]\in \bar\cG,[\m_2] \in \bar\cG_G : 
[\m_1]\circ [\m_2]=[\m]} f[\m_1] \xi[\m_2].
\end{equation}
As in the previous case, we define on the space $\bar\cH_G$ of
finitely supported functions $\xi:\bar\cG_G \to \C$ the inner product
\begin{equation}\label{innprodquot}
\langle \xi,\xi'\rangle =\sum_{[\m]} \overline{\xi[\m]}\xi'[\m]. 
\end{equation}
Once again we see that, in this representation, the adjoint does not
correspond to the involution $f^\vee$ but it is instead given by the
involution in the algebra of creation and annihilation operators
\begin{equation}\label{AMquot}
(A_{[\m]}\xi)[\m']=\left\{ \begin{array}{ll} \xi[\m''] & \text{if }\,
[\m']=[\m]\circ[\m''] \\[2mm]
0 & \text{otherwise} \end{array}\right.
\end{equation}
\begin{equation}\label{AMquotstar}
(A_{[\m]}^*\xi)[\m']=\left\{ \begin{array}{ll} \xi[\m\circ \m'] & \text{if the composition
is possible} \\[2mm]
0 & \text{otherwise.} \end{array}\right.
\end{equation}
Again we have $\rho_G(\delta_{[\m]})=A_{[\m]}$ so that the algebra generated by
the $A_{[\m]}$ is the same as the image of $\C[\bar\cG]$ in the representation $\rho_G$
and the algebra of the creation and annihilation operators $A_{[\m]}$ and $A_{[\m]}^*$
is the involutive algebra in $\cB(\bar\cH_G)$ generated by
$\C[\bar\cG]$. In fact, the same argument we used before shows that
$A_{[\m]}^*$ defined as in \eqref{AMquotstar} is the adjoint of
$A_{[\m]}$ in the inner product \eqref{innprodquot}. 

\smallskip

We then have the following result.

\begin{thm}\label{finmult}
The Hamiltonian $H=H^R_{G}$ generating the time evolution $\sigma^R_t$
in the representation \eqref{repCGbar} has discrete spectrum 
$$ \Spec(H)=\{ \log(n) \}_{n\in\N}, $$ with finite multiplicities 
\begin{equation}\label{NnK}
 1\leq N_n \leq \# \pi_3(B_n), 
\end{equation}
where $B_n$ is the classifying space for branched coverings of order $n$.
\end{thm}

\proof It was proven in \cite{Brand} that the $n$-fold branched
covering spaces of a manifold $\m$, up to cobordism of branched
coverings, are parameterized by the homotopy classes
\begin{equation}\label{BnM}
 B_n(\m)=[\m,B_n], 
\end{equation}
where the $B_n$ are classifying spaces. In particular, cobordism
equivalence classes of $n$-fold branched coverings of the 3-sphere are
classified by the homotopy group
\begin{equation}\label{BnS3}
 B_n(\s)=\pi_3(B_n).
\end{equation}
The rational homotopy type of the classifying spaces $B_n$ is computed
in \cite{Brand} in terms of the fibration
\begin{equation}\label{Kfibration}
K(\pi,j-1) \to \bigvee^{t-1} \Sigma K(\pi,j-1) \to \bigvee^t K(\pi,j),
\end{equation}
which holds for any abelian group $\pi$ and any positive integers
$t,j\geq 2$, with $\Sigma$ denoting the suspension. For the $B_n$ one finds
\begin{equation}\label{BnQ}
B_n \otimes \Q = \bigvee^{p(n)} K(\Q,4)
\end{equation}
with the fibration
\begin{equation}\label{BnQfib}
S^3 \otimes \Q \to \bigvee^{p(n)-1} S^4\times \Q \to  B_n \otimes \Q, 
\end{equation}
where $p(n)$ is the number of partitions of $n$. The rational homotopy
groups of $B_n$ are computed from the exact sequence of the fibration
\eqref{BnQfib} (see \cite{Brand}) and are of the form $\pi_k(B_n)
\otimes \Q =\Q^D$ with 
\begin{equation}\label{QpikBn}
D=\left\{\begin{array}{ll} p(n) & k=4 \\ Q(\frac{k-1}{3},p(n)-1) &
 k=1,4,10 \mod 12, \text{ with } k\neq 1,4 \\
Q(\frac{k-1}{3},p(n)-1)+Q(\frac{k-1}{6},p(n)-1) & k\equiv 7 \mod 12 \\
0 & \text{otherwise} \end{array}\right.
\end{equation}
where
$$ Q(a,b)=\frac{1}{a}\sum_{d|a} \mu(d) b^{a/d} $$
with $\mu(d)$ the M\"obius function.
The result \eqref{QpikBn} then implies that the homotopy groups $\pi_3(B_n)$
satisfy $\pi_3(B_n)\otimes \Q =0$. 

Moreover, in \cite{Brand} the classifying
spaces $B_n$ are constructed explicitly by fitting together the classifying
space $BO(2)$, that carries the information on the branch locus, with the
classifying space $BS_k$, for $S_k$ the group of permutations of $k$ elements.
For example, in the case of normalized simple coverings of \cite{BraBru}, the 
classifying space is a mapping cylinder $BO(2)\cup_{BD_k} BS_k$, with $D_k$ the 
dihedral group, over the maps induced by the inclusion $D_k \hookrightarrow O(2)$
as the subgroup leaving the set of k-th roots of unity globally invariant, and  
$D_k \to S_k$ giving the permutation action on the k-th roots of unity. In the
case of \cite{Brand} that we consider here, where more general branched coverings are 
considered, the explicit form of $B_k$ in terms of $BO(2)$ and $BS_k$ is more
complicated, as it also involves a union over partitions of $k$, which accounts for the 
different choices of branching indices, of data of disk bundles associated to
each partition. The skeleta of the classifying space have finitely generated 
homology in each degree, \ie they are spaces of finite type, and simply 
connected in the case of \cite{Brand}.
By a result of Serre it is known that, for simply connected spaces of finite type, 
the homotopy groups are also finitely generated (\cf also \S 0.a of \cite{Henn}).
The condition $\pi_3(B_n)\otimes \Q =0$ then implies that the groups 
$\pi_3(B_n)$ are finite for all $n$. 

By the same argument used in Lemma \ref{HamM}, the Hamiltonian $H=H_G^R$
generating the time evolution $\sigma^R_t$ in the representation \eqref{repCGbar} 
is of the form
\begin{equation}\label{HamCGbar}
(H \, \xi) [\m] = \log(n) \, \xi[\m],
\end{equation}
where $\m$ is a branched cover of $\s$ of order $n$ branched along
$E\supset G$, for the given embedded graph $G$ specifying the representation.
Thus, the multiplicity of the eigenvalue $\log(n)$ is the number of
cobordism classes $[\m]$ branched along an embedded graph containing $G$ as a subgraph. 
This number $N_n=N_n(G)$ is bounded by
$1 \leq N_n(G) \leq \# \pi_3(B_n)$.
\endproof

The result can be improved by considering, instead of the Brand
classifying spaces $B_n$ of branched coverings, the more refined
Tejada classifying spaces $B_n(\ell)$ introduced in \cite{Te},
\cite{BrTe}. In fact, the homotopy group $\pi_3(B_n)$ 
considered above parameterizes branched
cobordism classes of branched coverings where the branch loci are
embedded manifolds of codimension two. Since in each cobordism
class there are representatives with such branch loci (\cf the 
discussion in Section \ref{GtoKsect} below) we can work with $B_n$
and obtain the coarse estimate above. However, in our construction 
we are considering branch loci that are, more generally, embedded
graphs and not just links. Similarly, our cobordisms are branched
over 2-complexes, not just embedded surfaces. In this case, the
appropriate classifying spaces are the generalizations $B_n(\ell)$
of \cite{Te}, \cite{BrTe}. These are such that $B_n(2)=B_n$ and
$B_n(\ell)$, for $\ell >2$, allows for branched coverings and
cobordisms where the branch locus has strata of some codimension $2\leq r
\leq \ell$. We have then the following more refined result.

\begin{cor}\label{finmult2}
The multiplicity $N_n(G)$ of the eigenvalue $\log(n)$ of the
Hamiltonian $H_G$ satisfies the estimate
\begin{equation}\label{NGBnell}
1 \leq N_n(G) \leq \# \pi_3(B_n(4)).
\end{equation}
\end{cor}

\proof In our construction, we are considering branched coverings of
the 3-sphere with branch locus an embedded graph $E\supset G$, up to
branched cover cobordism, where the cobordisms are branched over a
2-complex. Thus, the branch locus $E$ has strata of codimension two 
and three and the branch locus for the cobordism has strata of
codimension two, three, and four. Thus, we can consider, instead of
the classifying space $B_n$, the more refined $B_n(4)$. The results of
\cite{BrTe} show that $\pi_3(B_n)\cong \pi_3(B_n(3))$, while there 
is a surjection $\pi_3(B_n(3))\to \pi_3(B_n(4))$, so that we have
$\# \pi_3(B_n(4)) \leq \# \pi_3(B_n)$. Thus, the same argument of Theorem
\ref{finmult} above, using cobordisms with stratified branch loci, 
gives the finer estimate \eqref{NGBnell} for the multiplicities.
\endproof

\smallskip

We can then consider the partition function for the Hamiltonian of the
time evolution \eqref{sigmatnmbar}. To stress the fact that we work in
the representation $\rho=\rho_G$ associated to the subsemigroupoid 
$\bar\cG_K$ for a given knot $K$, we write $H=H(G)$. We then have
\begin{equation}\label{ZetaK}
Z_G(\beta)= \Tr(e^{-\beta H}) = \sum_n \exp(-\beta\, \log(n))\,
N_n(G).
\end{equation}
Thus, the question of whether the summability condition 
$\Tr(e^{-\beta H})<\infty$ 
holds depends on an estimate of the asymptotic growth of the cardinalities
$\# \pi_3(B_n(4))$ for large $n\to \infty$, by the estimate
\begin{equation}\label{Zetabound}
\zeta(\beta)=\sum_n n^{-\beta}\leq Z_G(\beta) \leq
\sum_n \#\pi_3(B_n(4))\, n^{-\beta} .
\end{equation}
This corresponds to the question of studying a generating function 
for the numbers $\#\pi_3(B_n(4))$. We will not pursue this further
in the present paper but we hope to return to this question in 
future work.

\smallskip

Notice that there is evidence in the results of \cite{BraBru} in favor of
some strong constraints on the growth of the numbers $\# \pi_3(B_n)$ 
(hence of the $\#\pi_3(B_n(4))$), based on the periodicities 
along certain arithmetic progressions of the localizations at primes.

In fact, it is proved in \cite{BraBru} that, at least for the classifying 
spaces $BR_n$ of normalized simple branched coverings,
in the stable range $n >4$ and for any given prime $p$,
the localizations $\pi_3(BR_n)_{(p)}$ satisfy the periodicity
$$ \pi_3(BR_n)_{(p)} = \pi_3(BR_{n+2^{a+i+1} p^{b+j}})_{(p)}, $$
for $n=2^a p^b m$ with $(2,m)=(p,m)=1$. The number $2^i p^j$ is
determined by homotopy theoretic data as described in 
Proposition 11 of \cite{BraBru}. Thus, one can consider associated
zeta functions
\begin{equation}\label{Zetap}
Z_p(\beta)= \sum_n \# \pi_3(BR_n)_{(p)} n^{-\beta}.
\end{equation}

\smallskip

If a finite summability $\Tr(e^{-\beta H})<\infty$ holds
for sufficiently large $\beta >>0$,  
then one can recover invariants of embedded graphs as zero temperature
KMS functionals, by considering 
functionals of the Gibbs form
\begin{equation}\label{GibbsKMS}
\varphi_{G,\beta}(f)=\frac{\Tr (\rho_G(f) e^{-\beta
H})}{\Tr(e^{-\beta H})},  
\end{equation}
where, for instance, $f$ is taken to be an invariant of embedded
graphs in 3-manifolds 
and $f(\m):=f(\pi_G^{-1}(G))$, for $\pi_G:\m\to \s$ the branched
covering map. In this 
case, in the zero temperature limit, \ie for $\beta \to \infty$, the weak
limits of states of the form \eqref{GibbsKMS} would give back the invariant of 
embedded graphs in $\s$ in the form 
$$ \lim_{\beta\to\infty} \varphi_{K,\beta}(f)= f(\bU_G). $$

\smallskip

Notice that, to the purpose of studying KMS states for the algebra with time evolution, 
the convergence of the partition function $Z_G(\beta)$ is not needed, as KMS states 
need not necessarily be of the Gibbs form \eqref{GibbsKMS}, \cf \cite{Haag}.
However, it is still useful to
consider the question of the convergence of the partition function $Z_G(\beta)$, since
Gibbs states of the form \eqref{GibbsKMS} may have applications to constructing 
interesting zeta functions for embedded graphs $G\subset S^3$. 

\smallskip

For instance, suppose given an
invariant $F$ of cobordism classes of embedded graphs in $S^3$. Cobordism for embedded 
graphs can be defined, for connected graphs, as in \cite{Tan}, and in the multi-connected
case using the same basic relation (attaching a 1-handle) as in the case of links, as in
\cite{Hoso}. An example of such an invariant can be obtained, for instance, by considering the
collection of links $T(G)$ constructed in \cite{Kauf} as an invariant of an embedded graph $G$ 
and define a total linking number of $T(G)$ by adding the total linking numbers of all the
links in the collection. 

\smallskip

Given such an invariant $F$, one can then consider, for a set of representatives of the classes
$[M]\in \pi_3(B_n)$, the values $F(\pi_{G'}\pi_G^{-1}(G))$ and form the series
\begin{equation}\label{ZetaFG}
\sum_n \sum_{[M]\in \pi_3(B_n)} F(\pi_{G'}\pi_G^{-1}(G)) \, n^{-\beta},
\end{equation}
where the inner sum is over the classes $[M]\in \pi_3(B_n)$ such that $M$ is a branched cover of
$S^3$ branched along a graph $E\supset G$.
Similarly, one can form variations of this same concept based on the zeta functions \eqref{Zetap}.
When the function $F$ on the set of the $\{ \pi_{G'}\pi_G^{-1}(G) \}$ is either
bounded or of some growth $\sim n^\alpha$, then the convergence of $Z_G(\beta)$ (or
of the $Z_p(\beta)$ of \eqref{Zetap}) would ensure the convergence of \eqref{ZetaFG}. 
Obviously such zeta functions are very complicated objects, even for very simple graphs $G$ and
it would be difficult to compute them explicitly, but it would be interesting to see whether
some variant of this idea might have relevance in the context of spin networks, spin foams, 
and dynamical triangulations.

\medskip

Finally, notice that, while the Hamiltonian $H$ of the time evolution $\sigma_t^L$
has finite multiplicities in the spectrum after passing to the quotient by the equivalence 
relation of cobordism (similarly for $\sigma^R_t$), the infinitesimal generator for the
time evolution $\sigma_t =\sigma^L_t \sigma_{-t}^R$ still has infinite multiplicities.
In fact, the time evolution \eqref{sigmatnm} is generated by an unbounded operator $D$
that acts on a densely defined domain in $\cH_G$ by
\begin{equation}\label{Dsigma}
D \, \delta_\m = \log\left(\frac{n}{m}\right) \, \delta_\m,
\end{equation}
with $n$ and $m$ the multiplicities of the two covering maps, as above.
This operator is not a good physical Hamiltonian since is does not have
a lower bound on the spectrum. It has the following property.

\begin{lem}\label{Dqdirac}
The operator $D$ of \eqref{Dsigma} has bounded commutators $[D,a]$ with
the elements of the involutive algebra generated (algebraically) by the $A_{[\m]}$
and $A_{[\m]}^*$.
\end{lem}

\proof It suffices to check that the commutators 
$[D,A_{[\m]}]$ and $[D,A_{[\m]}^*]$ are bounded. We have
$$ [D,A_{[\m]}^*] \delta_{[\m']} = \left( \log\left(\frac{nn'}{mm'}\right) - \log\left(\frac{n'}{m'}\right) \right) 
\delta_{[\m\circ \m']} = \log\left(\frac{n}{m}\right) \delta_{[\m\circ \m']} . $$
The case of $[D,A_{[\m]}]$ is analogous.
\endproof

Notice, however, that $D$ fails to be a Dirac operator in the sense of spectral triples, because of the
infinite multiplicities of the eigenvalues.

\section{From graphs to knots}\label{GtoKsect}

The Alexander branched covering theorem is greatly refined by the 
Hilden--Montesinos theorem, which ensures that all closed oriented
3-manifolds can be realized as branched covers of the 3-sphere,
branched along a knot or a link (see \cite{Hil}, \cite{Mont}, 
\cf also \cite{pr}).

One can see how to pass from a branch locus that is a multi-connected graph to 
one that is a link or a knot in the following way, \cite{BeEd}.
One says that two branched coverings $\pi_0:\m \to \s$ and $\pi_1:\m \to \s$
are $b$-homotopic if there exists a homotopy $H_t: \m \to \s$ with $H_0=\pi_0$,
$H_1=\pi_1$ and $H_t$ a branched covering, for all $t\in [0,1]$, with branch
locus an embedded graph $G_t \subset \s$. 

The ``Alexander trick'' shows that two branched coverings of the 3-ball $D^3\to D^3$ 
that agree on the boundary $S^2=\partial D^3$ are $b$-homotopic. 
Using this trick, one can pass, by a $b$-homotopy, 
from an arbitary branched covering to one that is {\em simple}, namely where all the 
fibers consist of al least $n-1$ points, $n$ being the order of the covering. Simple
coverings are {\em generic}. The same argument shows (\cite{BeEd}, Corollary 6.6)
that any branched covering $\m \to \s$ is $b$-homotopic to one where the branch set 
is a link. 

We restrict to the case where the embedded graphs $G$ and $G'$ are knots $K$ and $K'$ and 
we consider geometric correspondences $\cC(K,K')$ modulo the equivalence relation of
$b$-homotopy. Namely, we say that two geometric correspondences $\m_1,\m_2 \in \cC(K,K')$ 
are $b$-homotopic if there exist two homotopies $\Theta_t$, $\Theta_t'$ relating the branched covering maps 
$$ \s \stackrel{\pi_{K,i}}{\longleftarrow} \m \stackrel{\pi_{K',i}}{\longrightarrow} \s. $$

Since we have the freedom to modify correspondences by $b$-homotopies, we can as well 
assume that the branch loci are links. Thus, we are considering geometric correspondences 
of the form 
\bq \label{2coversLink} 
K \subset L \subset \s
\stackrel{\pi_K} \longleftarrow \m \stackrel{\pi_{K'}}
\longrightarrow \s \supset L'\supset K' ,
\nq 
where the branch loci are links $L$ and $L'$, containing the knots $K$
and $K'$, respectively.  
Notice also that, if we are allowed to modify the coverings by
$b$-homotopy, we can arrange so that,  
in the composition $\m_1 \circ \m_2$, the branch loci $L\cup \pi_k \pi_1^{-1}(L_2')$ and
$L''\cup \pi_{K''} \pi_2^{-1}(L_1')$ are links in $\s$.

We denote by $[\m]_b$ the equivalence class of a geometric
correspondence under the equivalence 
relation of $b$-homotopy.
The equivalence relation of $b$-homotopy is a particular case of the relation of branched 
cover cobordism that we considered above. In fact, the homotopy $\Theta_t$ can be realized by
a branched covering map $\Theta: \m \times [0,1] \to \s \times [0,1]$,
branched along a 2-complex 
$S=\cup_{t\in [0,1]} G_t$ in $\s \times [0,1]$.
Thus, by the same argument used to prove the compatibility of the composition of 
geometric correspondences with the equivalence relation of cobordism, we obtain the 
compatibility of composition
\begin{equation}\label{bcomp}
[\m_1]_b \circ [\m_2]_b = [\m_1\circ \m_2]_b.
\end{equation}  
The $b$-homotopy is realized by the cobordims $(\m_1\circ \m_2)\times [0,1]$ 
with the branched covering maps $\hat \Theta= \Theta \circ P_1$ and
$\hat \Theta'=\Theta'\circ P_2$.  

While the knots $K$ and $K'$ are fixed in the construction of
$\cC(K,K')$, the other components of  
the links $L$ and $L'$, when we consider the correspondences up to
$b$-homotopy, are only determined up to  
knot cobordism with trivial linking numbers (\ie as classes in the
knot cobordism subgroup of the link 
cobordism group, see \cite{Hoso}).

To make the role of the link components more symmetric, it is then
more natural in this setting to 
consider a category where the objects are cobordism classes of knots
$[K]$, $[K']$ and where the 
morphisms are given by the $b$-homotopy classes of geometric
correspondences $\cC([K],[K'])_b$. 
 
The time evolution considered above still makes sense on the
corresponding semigroupoid ring, since the  
order of the branched cover is well defined on the $b$-equivalence
class and multiplicative under composition 
of morphisms.

\section{Convolution algebras and 2-semigroupoids}\label{2catSec}

In noncommutative geometry, it is customary to replace the operation
of taking the quotient by an equivalence relation by forming a suitable
convolution algebra of functions over the graph of the equivalence relation. 
This corresponds to replacing an equivalence relation by the corresponding 
groupoid and taking the convolution algebra of the groupoid, \cf \cite{Co94}.

In our setting, we can proceed in a similar way and, instead of
taking the quotient by the equivalence relation of cobordism of
branched cover, as we did above, keep the cobordisms explicitly and
work with a 2-category.

\begin{lem}\label{2category}
The data of embedded graphs in the 3-sphere, 3-dimensional geometric correspondences, 
and 4-dimensional branched cover cobordisms form a 2-category $\cG^2$.
\end{lem}

\proof
We already know that embedded graphs and geometric correspondences form
semigroupoid with associative composition
of morphisms given by the fibered product of geometric correspondences.
Suppose given geometric correspondences $\m_1$ $\m_2$ and $\m_3$ in 
$\cC(G,G')$, and suppose given cobordisms $W_1$ and $W_2$ with
$\partial W_1 = \m_1\cup -\m_2$ and  $\partial W_2 = \m_2 \cup -\m_3$.
As we have seen in Lemma \ref{equivrel}, for the transitive 
property of the equivalence relation, the gluing of cobordisms 
$W_1 \cup_{\m_2} W_2$ gives a cobordism between $\m_1$ and $\m_3$ 
and defines in this way a composition of 2-morphisms that has the 
right properties for being the vertical composition in the 2-category. 
Similarly, suppose given correspondences
$\m_1, \tilde \m_1 \in \cC(G,G')$, and $\m_2, \tilde \m_2 \in \cC(G',G'')$,
with cobordisms $W_1$ and $W_2$ with $\partial W_1 = \m_1\cup -\tilde \m_1$
and $\partial W_2 = \m_2 \cup -\tilde\m_2$. 
Again by the argument of Lemma \ref{equivrel}, we know that 
the fibered product $W_1 \circ W_2$ 
defines a cobordism between the compositions $\m_1\circ \m_2$ and 
$\tilde \m_1\circ \tilde\m_2$. This gives the horizontal composition of
2-morphisms. By the results of Lemma \ref{equivrel} and an argument
like that of Proposition \ref{assoc}, one sees that both the vertical 
and horizontal compositions of 2-morphisms are associative.
\endproof

In the following, we denote the compositions of 2-morphisms
by the notation
\begin{equation}\label{verthor}
\text{horizontal (fibered product):} \ \ W_1 \circ W_2 \ \ \  \ \
\text{vertical (gluing):} \ \  W_1 \bullet W_2.
\end{equation}

We obtain a convolution
algebra associated to the 2-semigroupoid described above.

Consider the space of complex 
valued functions with finite support 
\begin{equation}\label{functionsU}
f: \cU \to \C
\end{equation}
on the set 
$$ \cU =\cup_{\m_1,\m_2\in \cG} \cU_{(\m_1,\m_2)}, $$
of branched cover cobordisms
\begin{equation}\label{cU}
\cU_{(\m_1,\m_2)} =\{ W  \,|\, \m_1\stackrel{W}{\sim} \m_2 \},
\end{equation}
with
$$ S \subset \s\times I \stackrel{q}{\leftarrow} W \stackrel{q'}{\rightarrow} \s\times I 
\supset S', $$
where $\sim$ denotes the equivalence relation given by branched cover
cobordisms with $\partial W=\m_1 \cup -\m_2$, compatibly with the
branched cover structures as in \S \ref{CobordSec} above.

\smallskip  

As in the case of the sets $\cC(G,G')$ of geometric correspondences
discussed in \S \ref{CorrSetSec},
the collection $\cU_{(\m_1,\m_2)}$ of cobordisms 
can be identified with a set of branched covering data of a
representation theoretic nature. In fact, as a PL manifold, one such
cobordism $W$ can be specified by assigning a representation 
\begin{equation}\label{pi1WS}
\sigma_W: \pi_1((\s\times I)\smallsetminus S) \to S_n ,
\end{equation}
which determines a covering space on the complement of the branch
locus $S$. 

\smallskip

This space of functions \eqref{functionsU} can be made into an algebra
$\cA(\cG^2)$ with the associative convolution product of the form
\begin{equation}\label{prodAG2}
(f_1 \bullet f_2)(W) = \sum_{W=W_1\bullet W_2} f_1(W_1) f_2(W_2),
\end{equation}
which corresponds to the vertical composition of 2-morphisms, namely
the one given by the gluing of cobordisms. Similarly,  one also has on
$\cA(\cG^2)$ an associative product which corresponds to  
the horizontal composition of 2-morphisms, given by the fibered
product of cobordisms, of the form 
\begin{equation}\label{prodAG2hor}
(f_1 \circ f_2)(W) = \sum_{W=W_1\circ W_2} f_1(W_1) f_2(W_2).
\end{equation}

We also have an involution compatible with both the horizontal 
and vertical product structure. In fact, consider the two involutions 
on the cobordisms $W$
\begin{equation}\label{barW}
W \mapsto \bar W=-W, \ \ \ W\mapsto W^\vee,
\end{equation}
where the first is the orientation reversal, so that if $\partial
W=\m_1 \cup -\m_2$ then $\partial \bar W= \m_2 \cup -\m_1$, while the
second extends the involution $\m\mapsto \m^\vee$ and exchanges the
two branch covering maps, that is, if $W$ has covering maps
$$ S \subset \s\times I \stackrel{q}{\leftarrow} W
\stackrel{q'}{\rightarrow}
\s\times I \supset S' $$
then $W^\vee$ denotes the same 4-manifold but with covering maps
$$ S' \subset \s\times I \stackrel{q'}{\leftarrow} W\stackrel{q}{\rightarrow}
\s\times I \supset S. $$
We define an involution on the algebra $\cA(\cG^2)$ by setting
\begin{equation}\label{fdag}
f^\dag(W)=\overline{f}(\bar W^\vee)
\end{equation}

\begin{lem}\label{invdag}
The involution $f\mapsto f^\dag$ makes $\cA(\cG^2)$ into an involutive
algebra with respect to both the vertical and the horizontal product.
\end{lem}

\proof We have $(f^\dag)^\dag=f$ since the two involutions $W\mapsto
\bar W$ and $W\mapsto W^\vee$ commute. We also have
$(af_1+bf_2)^\dag=\bar a f_1^\dag + \bar b f_2^\dag$. For the two
product structures, we have 
$$ \begin{array}{lr} 
\bar W =\bar W_1 \circ \bar W_2 & \text{ for } W= W_1 \circ W_2 \\
W^\vee = W_1^\vee \bullet W_2^\vee & \text{ for } W=W_1\bullet W_2
\end{array} $$
which gives
$$ (f_1 \circ f_2)^\dag (W)= \sum_{\bar W^\vee =\bar W_1^\vee \circ
\bar W_2^\vee} \overline{f_1}(\bar W_1^\vee)\overline{f_2}(\bar
W_2^\vee) = (f_2^\dag \circ f_1^\dag)(W) $$
$$ (f_1 \bullet f_2)^\dag (W)= \sum_{\bar W^\vee =\bar W_1^\vee \bullet
\bar W_2^\vee} \overline{f_1}(\bar W_1^\vee)\overline{f_2}(\bar
W_2^\vee) = (f_2^\dag \bullet f_1^\dag)(W). $$
\endproof

\section{Vertical and horizontal time evolutions}\label{VertHorTsect}

We say that $\sigma_t$ is a {\em vertical time evolution} on $\cA(\cG^2)$ if
it is a 1-parameter group of automorphisms of $\cA(\cG^2)$ with
respect to the product structure given by the vertical composition of
2-morphisms as in \eqref{prodAG2}, namely
$$ \sigma_t(f_1 \bullet f_2)=\sigma_t(f_1)\bullet \sigma_t(f_2). $$
Similarly, a {\em horizontal time evolution} on $\cA(\cG^2)$ satisfies
$$ \sigma_t(f_1 \circ f_2)=\sigma_t(f_1)\circ \sigma_t(f_2). $$

We give some simple examples of one type or the other first and then we move on to more 
subtle examples.

\begin{lem}\label{torderW}
The time evolution by order of the coverings defined in
\eqref{sigmatnm} extends to a horizontal time evolution on
$\cA(\cG^2)$. 
\end{lem}

\proof This clearly follows by taking the order of the cobordisms as branched coverings
of $\s\times I$. It is not a time evolution with respect to the
vertical composition.
\endproof

\begin{lem}\label{tinex}
Any numerical invariant that satisfies an inclusion-exclusion principle 
\begin{equation}\label{inclexcl}
\chi(A\cup B)=\chi(A)+\chi(B)-\chi(A\cap B)
\end{equation}
defines a vertical time evolution by
\begin{equation}\label{vertexclincl}
\sigma_t(f)(W)= \exp(it (\chi(W)-\chi(\m_2))) f(W),
\end{equation}
for $\partial W=\m_1 \cup -\m_2$.
\end{lem}

\proof This also follows immediately by direct verification, since
$$ \sigma_t(f_1*f_2)(W)=e^{it(\chi(W)-\chi(\m_2))} \sum_{W=W_1\cup_\m W_2} f_1(W_1) f_2(W_2) $$
$$ = e^{it(\chi(W_1) + \chi(W_2) - \chi(\m) -\chi(\m_2))} \sum_{W=W_1\cup_\m W_2} f_1(W_1) f_2(W_2) $$
$$ =\sum_{W=W_1\cup_\m W_2} e^{it(\chi(W_1)- \chi(\m))} f_1(W_1) e^{it(\chi(W_2)-\chi(\m_2))} f_2(W_2)
= (\sigma_t(f_1)*\sigma_t(f_2))(W). $$
\endproof

In particular, the following are two simple examples of this type of time evolution.

\begin{exa}\label{Eulchar}
Setting $\chi(W)$ to be the Euler characteristic gives a time evolution as in \eqref{vertexclincl}.
Since the 4-dimensional volume of the boundary 3-manifold $\m$ is zero, also setting $\chi(W)=Vol(W)$ 
gives a time evolution.
\end{exa}

A more elaborate example of this type is given in \S \ref{GaugeSec} below.

\section{Vertical time evolution: Hartle--Hawking
gravity}\label{HawkingSec}

We describe here a first non-trivial example of a vertical time evolution, which
is related to the Hartle--Hawking formalism of Euclidean quantum
gravity \cite{HartleHawking}. 

The classical Euclidean action for gravity on a 4-manifold $W$ with
boundary is of the form
\begin{equation}\label{EuclGravS}
S(W,g)= -\frac{1}{16 \pi} \int_W R \, dv - \frac{1}{8\pi} \int_{\partial W}
K \, dv ,
\end{equation}
where $R$ is the scalar curvature and $K$ is the trace of the II
fundamental form. 

In the Hartle--Hawking approach to quantum gravity, the transition
amplitude between two 3-dimensional geometries $\m_1$ and $\m_2$,
endowed with Riemannian structures $g_{\m_1}$ and $g_{\m_2}$ is given
by 
\begin{equation}\label{transitionampl}
\langle (\m_1,g_1), (\m_2,g_2) \rangle = \int e^{iS(g)} D[g],
\end{equation}
in the Lorentzian signature, where the formal functional integration
on the right hand side involves also a summation over topologies,
meaning a sum over all cobordisms $W$ with $\partial W=\m_1 \cup
-\m_2$. In the Euclidean setting the probability amplitude $e^{iS(g)}$
is replaced by $e^{-S(g)}$, with $S(g)$ the Euclidean action
\eqref{EuclGravS}. We have suppressed the dependence of the
probability amplitude on a quantization parameter $\hbar$.

\smallskip

This suggests setting
\begin{equation}\label{HHtimev}
\sigma_t(f)(W,g) := e^{it S(W,g)} f(W,g),
\end{equation}
with $S(W,g)$ as in \eqref{EuclGravS}.
For \eqref{HHtimev} to define a vertical time evolution, \ie for it to satisfy
the compatibility
$\sigma_t(f_1 \bullet f_2)=\sigma_t(f_1)\bullet \sigma_t(f_2)$
with the vertical composition, we need to impose conditions on the
metrics $g$ on $W$ so that the gluing of the Riemannian data near the
boundary is possible when composing cobordisms $W_1\bullet W_2=W_1
\cup_\m W_2$ by gluing them along a common boundary $\m$. 

For instance, one can assume cylindrical metrics near the boundary, though
this is does not correspond to the physically interesting case of more
general space-like hypersurfaces. Also, one needs to restrict here to
cobordisms that are smooth manifolds, or to allow for weaker forms of
the Riemannian structure adapted to PL manifolds, as is done in the
context of Regge calculus of dynamical triangulations. 

Then, formally, one obtains states for this vertical time
evolution that can be expressed in the form of a functional
integration as
\begin{equation}\label{HHevolKMS}
\varphi_\beta(f)=\frac{\int f(W,g) e^{-\beta S(g)} D[g]}{\int
e^{-\beta S(g)} D[g]}.
\end{equation}

We give in the next section a more mathematically rigorous example of
vertical time evolution. 

\section{Vertical time evolution: index splitting and gauge moduli}\label{GaugeSec}

Consider again the vertical composition $W_1 \bullet W_2=W_1 \cup_{\m_2} W_2$ given 
by gluing two cobordisms along their common boundary. In order to
construct interesting time evolutions on the corresponding convolution
algebra, we consider the spectral theory of Dirac type operators on these
4-dimensional manifolds with boundary, \cf \cite{BleeBoo}.  

Consider first the simpler case where $X$ is a closed connected 4-manifold and
$\m$ is a hypersurface that partitions $X\smallsetminus M$ in two components
$X=X_1\cup_M X_2$ with boundary $\partial X_1 = M = -\partial X_2$. We assume that 
$X$ is endowed with a cylindrical metric on a collar neighborhood $M\times [-1,1]$
of the hypersurface $M$. Let $\cD$ be an elliptic differential operator on $X$ of 
Dirac type. We take it to be the Dirac operator assuming that $X$ is a spin $4$-manifold. 
The restriction $\cD|_{M\times [-1,1]}$ has the form 
$$ \cD|_{M\times [-1,1]}=c(\frac{\partial}{\partial s} + \cB), $$
where $c$ denotes Clifford multiplication by $ds$ and $\cB$ is the self-adjoint 
tangential Dirac operator on $M$. We let $P_{\geq}$ denote the spectral 
Atiyah--Patodi--Singer boundary conditions, \ie the projection onto the subspace of
the Hilbert space of square integrable spinors $L^2(M,\cS^+|M)$ spanned by the eigenvectors 
of $\cB$ with non-negative eigenvalues. Here $\cS=\cS^+\oplus \cS^-$ is the spinor bundle 
on $X$, with $\cD^+: C^\infty(X,\cS^+)\to C^\infty(X,\cS^-)$. The projection $P_{\leq}$ is 
defined similarly. Let $\cD_i$ denote the Dirac operator on $X_i$ with APS boundary conditions, 
that is, 
$$ \cD_1^+ : C^\infty(X_1,\cS^+, P_{\leq})\to C^\infty(X_1,\cS^-), \ \ \ 
\cD_2^+: C^\infty(X_1,\cS^+, P_{\geq})\to C^\infty(X_2,\cS^-), $$
where 
$$ \begin{array}{l}
C^\infty(X_1,\cS^+, P_{\leq}) = \{ \psi \in C^\infty(X_1,\cS^+)\,|\, P_{\leq}(\psi|_M)=0 \}, \\[2mm]
C^\infty(X_2,\cS^+, P_{\geq}) = \{ \psi \in C^\infty(X_2,\cS^+)\,|\, P_{\geq}(\psi|_M)=0 \}. \end{array} $$
The index of the Dirac operator $\cD$ is computed by the Atiyah--Singer index theorem
and is given by a local formula, while the index of $\cD_i$ is given by the Atiyah--Patodi--Singer
index theorem and consists of a local formula, together with a correction given by an eta invariant 
of the boundary manifold $M$. Moreover, one has the following splitting formula for the index
(\cf \cite{BleeBoo}, p.77)
\begin{equation}\label{splitInd}
\Ind(\cD) = \Ind(\cD_1) + \Ind(\cD_2) - \dim \Ker (\cB).
\end{equation}

In the case of 4-manifolds $W=W_1 \cup_M W_2$, where $\partial W= M_1 \cup -M_3$, $\partial W_1 = 
M_1 \cup -M_2$, and $W_2= M_2 \cup - M_3$, one can modify the above setting by imposing APS boundary 
conditions at both ends of the cobordims. Namely, we assume that $W$
is a smooth manifold with boundary endowed with a Riemannian metric with cylindrical ends 
$M_1\times [0,1]$ and $M_3 \times [-1,0]$, as well as a cylindrical metric on a collar neighborhood 
$M\times [-1,1]$.

Thus, the operator $\cD$ will be the Dirac operator with
APS boundary conditions $P_{\geq}$ and $P_{\leq}$ at $M_1$ and $M_3$, and similarly for the 
operators $\cD_1$ and $\cD_2$. We also denote by $\cB$, $\cB_1$ and $\cB_2$ the tangential Dirac
operators on $M$, $M_1$ and $M_2$, respectively. 
We then obtain a time evolution on the algebra $\cA(\cG^2)$ with the
product \eqref{prodAG2} associated to the splitting of the index, in the following way.

\begin{lem}\label{evolInd}
Let $W=W_1 \cup_M W_2$ be a composition of 4-dimensional cobordisms with metrics as above, 
and with $\cD$, $\cD_i$ the corresponding Dirac operators with APS boundary conditions. We let
\begin{equation}\label{deltaInd}
\delta(W):= \Ind(\cD)- \dim \Ker(\cB_2).
\end{equation}
Then setting
\begin{equation}\label{deltaIndevol}
\sigma_t(f) (W)= \exp(it \delta(W))\,\, f(W)
\end{equation}
defines a time evolution on $\cA(\cG^2)$ with the product
\eqref{prodAG2} of vertical composition.
\end{lem}

\proof Using the splitting formula \eqref{splitInd} for the index one
sees immediately that  
$$ \sigma_t(f_1 \bullet f_2)(W) =\sum_{W=W_1\bullet W_2}  
e^{it \delta(W)} f_1(W_1) f_2(W_2)  $$
$$ = \sum_{W=W_1\bullet W_2}  e^{it (\Ind \cD_1+\Ind \cD_2 - \dim\Ker\cB
-\dim \Ker \cB_2)} f_1(W_1) f_2(W_2)  $$ $$=\sum_{W=W_1\bullet W_2} 
e^{it \delta(W_1)} f_1(W_1) e^{it \delta(W_2)}  f_2(W_2)=
\sigma_t(f_1)\bullet \sigma_t(f_2) (W). $$ 
\endproof

\smallskip

The type of spectral problem described above arises typically in the
context of invariants of 4-dimensional geometries that
behave well under gluing.
A typical such setting is given by the topological quantum field
theories, as outlined in \cite{Atiyah}, where to every 3-dimensional
manifolds one assigns functorially a vector space and to every
cobordism between 3-manifolds a linear map between the vector spaces.

In the case of Yang--Mills gauge theory, the gluing theory for moduli
spaces of anti-self-dual $SO(3)$-connections on smooth 4-manifolds
(see \cite{TB} for an overview) shows that if $\m$ is a closed
oriented smooth 3-manifold that separates a closed smooth 4-manifold
$X$ in two connected pieces  
\begin{equation}\label{Xpm}
 X =X_+ \cup_{\m} X_- 
\end{equation}
glued along the common boundary $\m=\partial X_+ =-\partial X_-$, then
the moduli space $\cM(X)$ of gauge equivalence classes of framed
anti-self-dual $SO(3)$-connections on $X$ decomposes as a fibered
product 
\begin{equation}\label{fibermodspace}
 \cM(X) =\cM(X_+)\times_{\cM(\m)} \cM(X_-),
\end{equation}
where $\cM(X_\pm)$ are moduli spaces of anti-self-dual
$SO(3)$-connections on the 4-manifolds with boundary and $\cM(\m)$ is
a a neighborhood of the moduli space of gauge classes of flat 
connections on the 3-manifold $\m$. The fibered product is over 
the restriction maps induced by the inclusion of $\m$ in $X_\pm$.
Setting up the appropriate analytical theory to compute the virtual 
dimensions is a technically very demanding task a detailed discussion 
of which is beyond the scope of this short paper. We only mention
the fact that virtual dimensions are given by indices of elliptic
operators. These operators arise via deformation complexes 
$\Omega^0 \to \Omega^1 \to \Omega^2$,
where the forst map correspond to the infinitesimal gauge action
and the second to the linearization of the nonlinear elliptic equations
at a solution. The counting of virtual dimensions that corresponds to the 
fibered product formula \eqref{fibermodspace} of Donaldson--Floer theory
is given by a splitting formula for the index of the type discussed 
above in \eqref{splitInd}. One finds similar fibered product formulae
in the gluing theory of other gauge theoretic moduli spaces, such as
Seiberg--Witten (\cf \eg \cite{CW}). It would be interesting to see
if invariants of 4-manifolds constructed from various gauge theories
and topological quantum field theories can give rise to interesting
dynamics and equilibrium states on the algebra $\cA(\cG^2)$ of geometric
correspondences and cobordisms, in a way that uses more information
than just the virtual dimension of the moduli spaces.

\section{Horizontal time evolution: bivariant Chern
character}\label{KKchernSec} 

The time evolution of Lemma \ref{evolInd}, however, does not
detect the structure of $W$ as a branched cover of $\s\times I$
branched along an embedded surface $S \subset \s\times I$. Thus, there
is no reason why a time evolution defined in this way should also be
compatible with the other product given by the horizontal composition
of 2-morphisms. However, the time evolution \eqref{deltaIndevol} obtained 
using the splitting formula \eqref{splitInd} for the index suggests a
possible way to define other time evolutions, also related to
properties of an index, which would be compatible with the
horizontal composition.

\smallskip

Although we are working here in the commutative context, in view of
the extension to noncommutative spectral correspondences outlined in
the next section, we give here a formulation using the language of 
KK-theory and cyclic cohomology that carries over naturally to the
noncommutative cases. 

\smallskip

In noncommutative geometry, one thinks of the index theorem as a
pairing of K-theory and K-homology, or equivalently as the pairing
$\langle ch_n(e), ch_n(x)\rangle$ of
Connes--Chern characters
\begin{equation}\label{CCchar}
ch_n: K_i(\cA)\to HC_{2n+i}(\cA) \ \ \text{ and } \ \ ch_n: K^i(\cA)\to
HC^{2n+i}(\cA),
\end{equation}
under the natural pairing of cyclic homology and cohomology, \cf \cite{Co94}.

\smallskip

Recall that cyclic (co)homology has a natural description in terms of
the derived functors $\Ext$ and $\Tor$ in the abelian category of
cyclic modules (\cf \cite{Co-ext}), namely 
\begin{equation}\label{ExtTor}
HC^n(\cA)= \Ext^n_\Lambda(\cA^\natural,\C^\natural) \ \ \text{ and } \ \
HC_n(\cA)= \Tor_n^\Lambda(\C^\natural,\cA^\natural),
\end{equation}
where $\Lambda$ denotes the cyclic category and $\cA^\natural$ is the
cyclic module associated to an associative algebra $\cA$.

\smallskip

It was shown in \cite{Nistor} that the characters \eqref{CCchar}
extend to a bivariant Connes--Chern character
\begin{equation}\label{bivariantCC}
ch_n: KK^i(\cA,\cB) \to \Ext^{2n+i}_\Lambda
(\cA^\natural,\cB^\natural)
\end{equation}
defined on KK-theory, with the natural cap product pairings
\begin{equation}\label{TorExtpair}
\Tor^\Lambda_m(\C^\natural,\cA^\natural)\otimes
\Ext^n_\Lambda(\cA^\natural, \cB^\natural) \to 
\Tor^\Lambda_{m-n}(\C^\natural,\cB^\natural)
\end{equation}
corresponding to an index theorem
\begin{equation}\label{bivind}
\psi = ch(x) \phi, \ \ \text{ with } \ \  \phi(e\circ x)=\psi(e).
\end{equation}
The construction of a bivariant Connes--Chern character that is 
fully compatible with the composition products, namely that sends 
the Kasparov product
$$ \circ: KK^i(\cA,\cC)\times KK^j (\cC,\cB)\to KK^{i+j}(\cA,\cB) $$ 
to the Yoneda products 
\begin{equation}\label{yoneda}
\Ext^{2n+i}_\Lambda(\cA^\natural,\cC^\natural)\times 
\Ext^{2m+j}_\Lambda (\cC^\natural,\cB^\natural) 
\to \Ext^{2(n+m)+i+j}(\cA^\natural,\cB^\natural),
\end{equation}
requires a modification of both $KK$-theory and cyclic cohomology. 
Such a general form of the bivariant Connes--Chern character is 
given in \cite{Cu}.

\smallskip

The construction of \cite{CoSka} of geometric correspondences
realizing KK-theory classes shows that, given manifolds $X_1$ and
$X_2$, classes in $KK(X_1,X_2)$ are realized by geoemtric data $(Z,E)$
of a manifold $Z$ with submersions $X_1 \leftarrow Z \rightarrow X_2$
and a vector bundle $E$ on $Z$. The Kasparov product 
$x \circ y \in KK(X_1,X_3)$, for $x=kk(Z,E)\in KK(X_1,X_2)$ and
$y=kk(Z',E')\in KK(X_2,X_3)$, is given by the fibered product
$x\circ y = kk(Z\circ Z',E\circ E')$, where
$$ Z\circ Z'=Z\times_{X_2} Z' \ \ \text{ and } \ \ 
E\circ E'= \pi_1^* E \times \pi_2^* E'. $$ 

\smallskip

To avoid momentarily the complication caused by working with manifolds
with boundary, we consider the simpler situation where $W$ is a
4-manifold endowed with branched covering maps to a closed 4-manifold
$X$ (for instance $\s\times S^1$ or $S^4$) instead of $\s \times [0,1]$,
\begin{equation}\label{S31coverings}
 S \subset X \stackrel{q}{\longleftarrow} W
\stackrel{q'}{\longrightarrow} X \supset S' 
\end{equation}
branched along surfaces $S$ and $S'$ in $X$.

We can then think of an elliptic operator $\cD_W$ on a 4-manifold $W$,
which has branched covering maps as in \eqref{S31coverings},
as defining an unbounded Kasparov bimodule, \ie as defining a KK-class
$[\cD_W] \in KK(X,X)$. We can think of this class as being
realized by a geometric correspondence in the sense of \cite{CoSka}
$$ [\cD_W]= kk(W,E_W), $$
with the property that, for the horizontal composition $W=W_1\circ
W_2=W_1\times_X W_2$ we have
$$ [\cD_{W_1}]\circ [\cD_{W_2}] = kk(W_1,E_{W_1})\circ
kk(W_2,E_{W_2})= kk(W,E_W) = [\cD_W]. $$

\smallskip

Under a bivariant Chern character that is compatible with the 
composition products these classes will map to elements in the
Yoneda algebra 
\begin{equation}\label{yonedaalg}
ch_n([\cD_W]) \in \cY:=\oplus_j \Ext^{2n+j}(\cA^\natural,\cA^\natural)
\end{equation}
$$ ch_n([\cD_{W_1}]) ch_m([\cD_{W_2}])= ch_{n+m}([\cD_{W_1}]\circ
[\cD_{W_2}]). $$
Let $\chi:\cY \to \C$ be a character of the Yoneda algebra.
Then by composing $\chi \circ ch$ we obtain 
$$ \chi ch([\cD_{W_1}]\circ [\cD_{W_2}]) = \chi ch([\cD_{W_1}]) \chi
ch([\cD_{W_2}]) \in \C. $$
This can be used to define a time evolution for the horizontal product
of the form
$$ \sigma_t(f)(W)= |\chi ch ([\cD_W])|^{it} \, f(W) . $$

\section{Noncommutative spaces and spectral correspondences}\label{NCGsec}

We return now briefly to the problem of spectral correspondences
of \cite{co1}, mentioned in the introduction.

Recall that a spectral triple $(\cA,\cH,D)$ consists of the data of
a unital involutive algebra $\cA$, a representation $\rho: \cA \to \cB(\cH)$
as bounded operators on a Hilbert space $\cH$ and a self-adjoint operator
$D$ on $\cH$ with compact resolvent, such that $[D,\rho(a)]$ is a
bounded operator for all $a\in \cA$. We extend this notion to a 
correspondence in the following way, following \cite{co1}.

\begin{defn}\label{SpCorr}
A spectral correspondence is a set of data $(\cA_1,\cA_2, \cH, D)$,
where $\cA_1$ and $\cA_2$ are unital involutive algebras, with 
representations $\rho_i:\cA_i \to \cB(\cH)$, $i=1,2$, as bounded 
operators on a Hilbert space $\cH$, such that
\begin{equation}\label{comm12}
[\rho_1(a_1),\rho_2(a_2)]=0, \ \ \  \forall a_1\in\cA_1, \ 
\forall a_2\in \cA_2,
\end{equation}
and with a self-adjoint operator $D$ with compact resolvent, such
that
\begin{equation}\label{Dcomm12}
[[D,\rho_1(a_1)],\rho_2(a_2)]=0, \ \ \  \forall a_1\in\cA_1, \ 
\forall a_2\in \cA_2,
\end{equation}
and such that $[D,\rho_1(a_1)]$ and $[D,\rho_2(a_2)]$ are bounded
operators for all $a_1\in\cA_1$ and $a_2\in\cA_2$. A spectral correspondence 
is even if there exists an operator $\gamma$ on $\cH$ with $\gamma^2=1$ and
such that $D$ anticommutes with $\gamma$ 
and $[\gamma,\rho_i(a_i)]=0$ for all $a_i\in\cA_i$,
$i=1,2$. A spectral correspondence is odd if it is not even.
\end{defn}

One might relax the condition of compact resolvent on the operator $D$ 
if one wants to allow more degenerate types of operators in the 
correspondences, including possibly $D\equiv 0$, as seems desirable
in view of the considerations of \cite{co1}. For our purposes here,
we consider this more restrictive definition. Notice also that
the condition \eqref{Dcomm12} also implies $[[D,\rho_2(a_2)],\rho_1(a_1)]=0$
because of \eqref{comm12}. 

A more refined notion of spectral correspondences as morphisms between 
spectral triples, in a setting for families, is being developed by B.~Mesland, \cite{Mesland}. 

\smallskip

We first show that our geometric correspondences define commutative
spectral correspondences and then we give a noncommutative example
based on taking products with finite geometries as in \cite{co1}.

\begin{lem}\label{commSpCorr}
Suppose given a closed connected oriented smooth 3-manifold 
with two branched covering maps 
$\s \stackrel{\pi_1} \longleftarrow \m \stackrel{\pi_2}
\longrightarrow \s$.
Given a choice of a Riemannian metric and a spin structure 
on $\m$, this defines a spectral correspondence for 
$\cA_1=\cA_2=C^\infty(\s)$. 
\end{lem}

\proof We consider the Hilbert space $\cH =L^2(\m,S)$, where $S$ 
is the spinor bundle on $\m$ for the chosen spin structure. Let 
$\dirac_\m$ be the corresponding Dirac operator. The covering
maps $\pi_i$, for $i=1,2$, determine representations $\rho_i: C^\infty(\s)
\to \cB(\cH)$, by $\rho_i(f)=c(f\circ \pi_i)$, where $c$ denotes the
usual action of $C^\infty(\m)$ on $\cH$ by Clifford multiplication on spinors. 
Then we have $[\dirac_\m,\rho_i(f)]=c(d(f\circ \pi_i))$, 
which is a bounded operator on $\cH$. All the commutativity conditions 
are satisfied in this case.
\endproof

Let $\cA$ and $\cB$ be finite dimensional unital (noncommutative) 
involutive algebras. 
Let $V$ be a finite dimensional vector space with commuting actions 
of $\cA$ and $\cB$. Let $T \in \End(V)$ be a
linear map such that $[[T,a],b]=0$ for all $a\in \cA$ and $b\in \cB$.
Then we obtain noncommutative spectral correspondences of the type
described in the last section of \cite{co1} in the following way.

\begin{lem}\label{NCspcorr}
The cup product $S_\m \cup S_F$ of 
$S_\m=(C^\infty(\s),C^\infty(\s),L^2(\m,S),\dirac_\m)$ and $S_F=(A,B,V,T)$ 
defines a noncommutative 
spectral correspondence for the algebras $C^\infty(\s)\otimes \cA$ and
$C^\infty(\s)\otimes \cB$. 
\end{lem}

\proof We simply adapt the usual notion of cup product for spectral 
triples to the case of correspondences. If the correspondence 
$(A,B,V,T)$ is even, with grading $\gamma$, then we consider the
Hilbert space $\cH=L^2(\m,S)\otimes V$ and the operator 
$D= T\otimes 1 + \gamma \otimes \dirac_\m$. Then the usual argument
for cup products of spectral triples show that $(C^\infty(\s)\otimes A,
C^\infty(\s)\otimes B, \cH, D)$ is an odd spectral correspondence. 
Similarly, if $(A,B,V,T)$ is odd, then take $\cH =L^2(\m,S)\otimes V
\oplus L^2(\m,S)\otimes V$, with the diagonal actions of 
$C^\infty(\s)\otimes \cA$ and $C^\infty(\s)\otimes \cB$. Consider then
the operator
$$ D= \left(\begin{matrix} 0 & \delta^* \\ \delta & 0 \end{matrix}\right), $$
for $\delta = T\otimes 1 + i \otimes \dirac_\m$. Then, by the same
standard argument that holds for spectral triples, the data
$(C^\infty(\s)\otimes A, C^\infty(\s)\otimes B, \cH, D)$ form an even
spectral correspondence with respect to the $\Z/2\Z$ grading
$$ \gamma = \left(\begin{matrix} 1 & 0 \\ 0 & -1 \end{matrix}\right). $$
In either case, we denote the resulting correspondence 
$(C^\infty(\s)\otimes A, C^\infty(\s)\otimes B, \cH, D)$
as the cup product $S_\m \cup S_F$.
\endproof

We can then form a convolution algebra on the space of correspondences, 
using the equivalence relation given by cobordism of branched 
covering spaces of \S \ref{CobordSec} above, as in \S \ref{2catSec} above.
This requires extending the equivalence relation defined by cobordisms
of branched coverings to the case of the product by a finite geometry. 
We propose the following construction.

The existence of a cobordism $W$ of branched coverings between two
geometric correspondences $\m_1$ and $\m_2$ in $\cC(K,K')$ implies the
existence of a spectral correspondence {\em with boundary} of the form
$$ S_W = (C^\infty(\m_1),C^\infty(\m_2),L^2(W,S),\dirac_W). $$
We will not discuss here the setting of spectral triples with boundary.
A satisfactory theory was recently developed by Chamseddine and Connes
(\cf \cite{CC}). We only recall here briefly the following notions, from
\cite{Co}. A spectral triple with boundary $(\cA,\cH,D)$ is {\em boundary 
even} if there is a $\Z/2\Z$-grading $\gamma$ on $\cH$ such that
$[a,\gamma]=0$ for all $a\in \cA$ and $Dom(D)\cap \gamma Dom(D)$ is
dense in $\cH$. The boundary algebra $\partial\cA$ is the quotient
$\cA/(J\cap J^*)$ by the two-sided
ideal $J=\{a\in\cA|a Dom(D)\subset \gamma Dom(D)\}$. The boundary
Hilbert space $\partial\cH$ is the closure in $\cH$ of $D^{-1}Ker D_0^*$,
where $D_0$ is the symmetric operator obtained by restricting 
$D$ to $Dom(D)\cap \gamma Dom(D)$. The boundary algebra acts on the
boundary Hilbert space by $a-D^{-2}[D^2,a]$. The boundary Dirac operator
$\partial D$ is defined on $D^{-1}Ker D_0^*$ and satisfies 
$\langle\xi,\partial D\eta\rangle=\langle\xi,D\eta\rangle$ for
$\xi\in \partial\cH$ and $\eta\in D^{-1}Ker D_0^*$. It has bounded
commutators with $\partial\cA$. 

One can extend from spectral triples to correspondences, by
having two commuting representations of $\cA_1$ and $\cA_2$ on
$\cH$ with the properties above and such that the resulting
boundary data
$(\partial\cA_1,\partial\cA_2,\partial\cH,\partial D)$ define
a spectral correspondence.  

If one wants to extend to the product geometries the condition of cobordism
of geometric correspondences, it seems that one is inavitably faced
with the problem of defining spectral triples with corners. In fact,
if $S_W$ and $S_F$ are both spectral triples with boundary, then their
cup product $S_W \cup S_F$ would not longer give rise to a spectral
triple with boundary but to one with corners. At present there isn't a
well defined theory of spectral triples with corners. However, we can
still propose a way of dealing with products of cobordisms by finite
noncommutative geometries, which remains within the theory of spectral
triples with boundary. To this purpose, we assume that the finite part
$S_F$ is an ordinary spectral triple, while only the cobordims part is
a spectral triple with boundary. We then relate the cup product 
$S_W \cup S_F$ to the spectral correspondences $S_{\m_i}\cup S_{F_i}$
via the boundary $\partial S_W$ and bimodules relating the $S_{F_i}$
to $S_F$. More precisely, we consider the following data.

Suppose given $\m_i \in \cC(K,K')$, $i=1,2$ as above and finite spectral correspondences
$S_{F_i}=(A_i,B_i,V_i,T_i)$.  
Then we say that the cup products $S_{\m_i}\cup S_{F_i}$
are related by a spectral cobordism if the following conditions hold.
The geometric correspondences are equivalent $\m_1 \circ \m_2$ via a 
cobordism $W$. There exist finite dimensional 
(noncommutative) algebras $R_i$, $i=1,2$ together with $R_i$--$A_i$  
bimodules $E_i$ and $B_i$--$R_i$ bimodules $F_i$, with connections. 
There exists
a finite spectral correspondence $S_F=(R_1,R_2,V_F,D_F)$ such that
$S_W \cup S_F =(\cA,\cH,\cD)$ is a spectral triple with boundary in the 
sense of Chamseddine--Connes with 
$$ \partial\cA = \oplus_{i=1,2} C^\infty(\m_i)\otimes R_i  $$
$$ \partial\cH = \oplus_{i=1,2} L^2(\m_i,S)\oplus 
(E_1\otimes_{A_1} V_i \otimes_{B_i} F_i) $$
and $\partial\cD$ gives the cup product of the
Dirac operators $\partial_{\m_i}$ with the $T_i$, with the latter twisted
by the connections on $E_i$ and $F_i$.

We do not give more details here. In fact, in order to use this
notion to extend the equivalence relation of cobordims of branched
coverings and the 2-category we considered in \S \ref{2catSec} above
to the noncommutative case, one needs a gluing theory for spectral
triples with boundary that makes it possible to define the
horizontal and vertical compositions of 2-morphisms as in the
case of $W_1\circ W_2$ and $W_1 \bullet W_2$. The analysis necessary
to develop such gluing results is beyond the scope of this paper
and the problem will be considered elsewhere.

\section{Questions and future work}\label{FutSec}

We sketch briefly an outline of ongoing work where the construction
presented in this paper is applied to other constructions related to 
noncommutative geometry and knot invariants.  

\subsection{Time evolutions and moduli spaces}

We have constructed vertical time evolutions from virtual dimensions
of moduli spaces. It would be more interesting to construct time
evolutions on the algebra of correspondences, in such a way that
the actual gauge theoretic invariants obtained by integrating certain
differential forms over the moduli spaces can be recovered as
low temperature equilibrium states. The formal path integral
formulations of gauge theoretic invariants of 4-manifolds suggests
that something of this sort may be possible, by analogy to the case we
described of Hartle--Hawking gravity. In the case of the horizontal
time evolution, it would be interesting to see if that can also be
related to gauge theoretic invariants. The closest model available
would be the gauge theory on embedded surfaces developed in \cite{KM}.

\subsection{Categorification and homology invariants}

We have constructed a category of knots and links, or more generally of embedded graphs,
where it is possible to use homological algebra to construct complexes and cohomological
invariants. The process of categorifications in knot theory, applied
to a different category of knots, has already proved very successful
in deriving new knot invariants such as Khovanov homology. We intend 
to investigate possible constructions of cohomological invariants using
the category defined in this paper.

\subsection{Noncommutative spaces and dynamical systems}

Another way to construct noncommutative spaces out of the geometric
correspondences considered here is via the subshifts of finite type
constructed in \cite{SW} out of the representations 
$\sigma: \pi_1(\s\smallsetminus L) \to S_m$ describing branched coverings.
A subshift of finite type naturally determines a noncommutative space
in the form of associated Cuntz--Krieger algebras. The covering 
moves (or colored Reidemeister moves) of \cite{Pier} will determine 
correspondences between these noncommutative spaces.

\end{document}